# Direct observation of nanoscale pinning centers in Ce(Co$_{0.8}$Cu$_{0.2}$)$_{5.4}$ permanent magnets

Nikita Polin[1], Shangbin Shen[2], Fernando Maccari[2], Alex Aubert[2], Esmaeil Adabifiroozjaei[3], Tatiana Smoliarova[5], Yangyiwei Yang[2], Xinren Chen[1], Yurii Skourski[6], Alaukik Saxena[1], András Kovács[7], Rafal E. Dunin-Borkowski[7], Michael Farle[5], Bai-Xiang Xu[2], Leopoldo Molina-Luna[3], Oliver Gutfleisch[2], Baptiste Gault[1,4], Konstantin Skokov[2]

[1] Max-Planck-Institut für Nachhaltige Materialien, 40237 Düsseldorf, Germany
[2] Institute of Materials Science, Technische Universität Darmstadt, 64287 Darmstadt, Germany
[3] Advanced Electron Microscopy Division, Institute of Materials Science, Department of Materials and Geosciences, Technische Universität Darmstadt, Peter-Grünberg-Str. 2, Darmstadt 64287, Germany
[4] Department of Materials, Royal School of Mines, Imperial College, Prince Consort Road, London SW7 2BP, United Kingdom
[5] Faculty of Physics and Center for Nanointegration (CENIDE), University Duisburg Essen, Duisburg 47057, Germany
[6] Dresden High Magnetic Field Laboratory (HLD-EMFL), Helmholtz-Zentrum Dresden-Rossendorf, Dresden 01328, Germany
[7] Ernst Ruska-Centre for Microscopy and Spectroscopy with Electrons, Forschungszentrum Jülich, Jülich 52425, Germany

Keywords: pinning, APT, CeCo$_5$, micromagnetics, nanostructure, electron microscopy, giant intrinsic magnetic hardness

# Graphical abstract

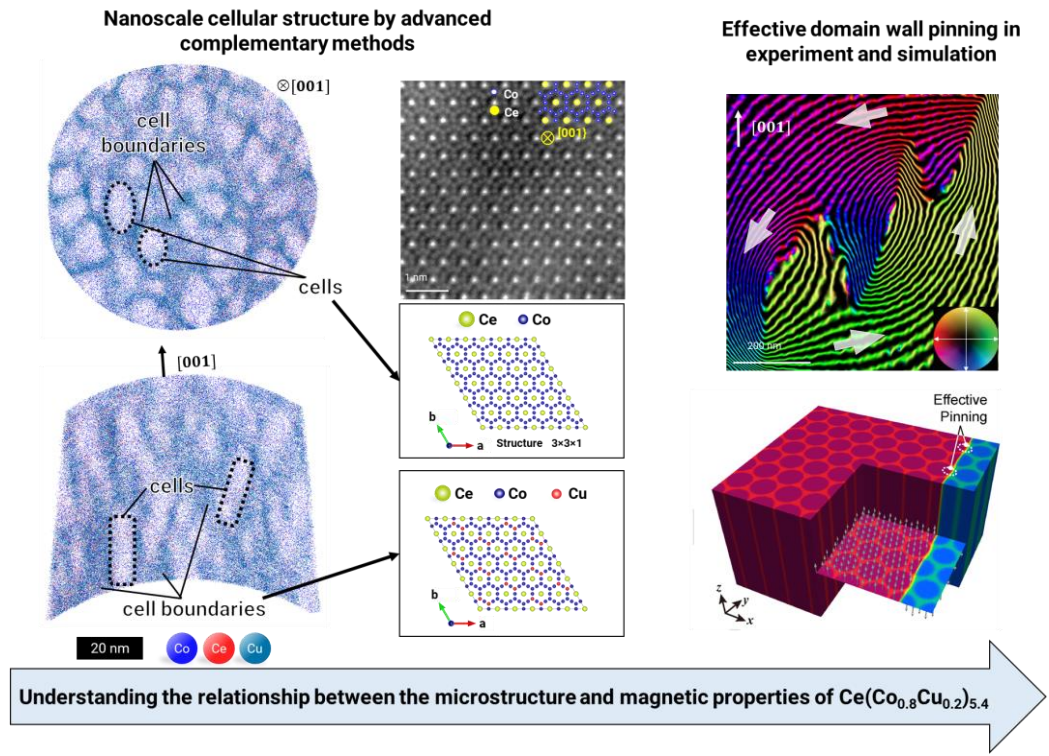

# Abstract

Permanent magnets containing rare earth elements are essential components for the electrification of society. Ce(Co$_{1-x}$Cu$_x$)$_5$ permanent magnets are a model system known for their substantial coercivity, yet the underlying mechanism remains unclear. Here, we investigate Ce(Co$_{0.8}$Cu$_{0.2}$)$_{5.4}$ magnets with a coercivity of ~1 T. Using transmission electron microscopy (TEM) and atom probe tomography (APT), we identify a nanoscale cellular structure formed by spinodal decomposition. Cu-poor cylindrical cells (~5-10 nm in diameter, ~20 nm long) have a disordered CeCo$_5$-type structure and a composition Ce(Co$_{0.9}$Cu$_{0.1}$)$_{5.3}$. Cu-rich cell boundaries are ~ 5 nm thick and exhibit a modified CeCo$_5$ structure, with Cu ordered on the Co sites and a composition Ce(Co$_{0.7}$Cu$_{0.3}$)$_{5.0}$. Micromagnetic simulations demonstrate that the intrinsic Cu concentration gradients up to 12 at.% Cu/nm lead to a spatial variation in magnetocrystalline anisotropy and domain wall energy, resulting in effective pinning and high coercivity. Compared to Sm$_2$Co$_{17}$-type magnets, Ce(Co$_{0.8}$Cu$_{0.2}$)$_{5.4}$ displays a finer-scale variation of conventional pinning with lower structural and chemical contrast in its underlying nanostructure. The identification of nanoscale chemical segregation in nearly single-phase Ce(Co$_{0.8}$Cu$_{0.2}$)$_{5.4}$ magnets provides a microstructural basis for the long-standing phenomenon of "giant intrinsic magnetic hardness" in systems such as SmCo$_{5-x}$M$_x$, highlighting avenues for designing rare-earth-lean permanent magnets via controlled nanoscale segregation.

# 1 Introduction

Permanent magnets containing rare earth elements are essential components in high-efficiency electric motors and energy conversion [1], which are crucial for the widespread electrification of transportation. Enhancing the performance of these magnets is a key technological challenge in the global effort to achieve net-zero carbon emissions [2]. Among the earliest rare-earth-based permanent magnets, $SmCo_5$ and $CeCo_5$ compounds were notable for demonstrating a significantly higher maximum energy product $BH_{\max}$ compared to earlier systems such as AlNiCo magnets [3]. Today, $SmCo_5$-based materials serve as model systems for studying nucleation-type magnets [4–8], and, industrially, the Cu-enriched $SmCo_5$ (1:5) phase plays a pivotal role in the development of $Sm_2Co_{17}$-type magnets with the composition $Sm(Co,Fe,Cu,Zr)_{7\pm\delta}$, acting as a pinning site for domain boundaries and thereby contributing to the high coercivity of these materials [9–13].

Compared to Sm, Ce is more abundant, less expensive, and considered less critical from both economic and environmental perspectives [14]. As a result, Ce-containing permanent magnets are a promising alternative for applications that require a balance between performance and sustainability. These materials can bridge the gap between high-end SmCo- and NdFeB-based magnets and lower-performance systems such as AlNiCo. Binary $SmCo_5$ magnets are classified as nucleation-type magnets [8], crystallizing in the hexagonal $CaCu_5$-type structure (space group P6/mmm) [15,16]. Upon substitution of Co with Cu (up to ~20 at.%), the ternary compounds $Sm(Co_{1-x}Cu_x)_5$ and $Ce(Co_{1-x}Cu_x)_5$ exhibit significant coercivity, reaching ~1 T after the proper heat treatment procedure known as ageing [3].

In the 1970s, it was concluded that $Sm(Co_{1-x}Cu_x)_5$ and related systems like $YCo_{5-x}Ni_x$, $ThCo_{5-x}Ni$, $SmCo_{3-x}Ni$, $SmCo_{2-x}Ni_x$, $Sm_2Co_{17-x}Al_x$ and others exhibit the so called "giant intrinsic magnetic hardness", fundamentally distinct from fine-particle or precipitation-hardened magnets [17–20]. However, the origin of this phenomenon remained unclear. Around the same time, Leamy *et al.* [21], using electron diffraction in transmission electron microscopy (TEM), reported nanoscale precipitates with a $Ce_2Co_{17}$-type phase in $CeCo_5$ magnets, suggesting a pinning-type coercivity mechanism. However, no compositional data for these precipitates were provided. To our knowledge, these findings have not been independently reproduced or verified in the subsequent literature.

Later, Girodin *et al.* [22], studied a range of compositions from $CeCo_5$ to $CeCu_5$, and identified a solid-state miscibility gap, in which two $CaCu_5$-type phases - one Co-rich and the other Cu-rich - coexist in thermodynamic equilibrium. This miscibility gap, with a critical temperature slightly above 800°C, promotes phase separation and facilitates the formation of microstructural features that act as effective domain wall pinning centers, enhancing coercivity. However, increasing Cu content concurrently lowers the saturation magnetization and the Curie temperature of the system [3,22], which can drop below room temperature, when the Cu concentration exceeds that of Co.

More recently, studies have shown that in Fe- and Cu-doped Ce(Co,Fe,Cu)$_5$ systems, dopants solubility significantly affects the thermal stability of the 1:5 matrix phase [23]. Low-temperature annealing leads to decomposition into 2:7 and 5:19 phases. Additionally, stacking faults in heat-treated samples have been proposed as a key contributor to the high coercivity. Further work [24] has shown that off-stoichiometric Ce(CoCu)$_{5+\delta}$ sintered magnets exhibit improved magnetic performance over stoichiometric counterparts, emphasizing the importance of precise compositional control to optimize magnetic properties.

To address the uncertainties surrounding the nanostructural origin of the coercivity, we synthesized off-stoichiometric Ce(Co$_{0.8}$Cu$_{0.2}$)$_{5.4}$ crystals, optimizing both composition and heat treatment to achieve a high coercivity of approximately 1 T. We refer to these samples with the nominal composition Ce(Co$_{0.8}$Cu$_{0.2}$)$_{5.4}$. We investigated the microstructure, nanoscale structure (referred to as nanostructure in the following) and magnetic domain structure of these magnets using state-of-the-art characterization techniques across different length scales, including magnetic force microscopy (MFM), Fresnel defocus imaging and off-axis electron holography (EH) in Lorentz TEM, and atom probe tomography (APT). Our analyses reveal a distinct cellular structure composed of elongated, Cu-poor cylindrical cells (~5-10 nm in diameter and ~20 nm in length, ~6 at.% Cu below average), enclosed by ~ 5 nm thick Cu-rich cell boundaries (~6 at.% Cu above average). Contrary to Leamy *et al*. [21], we found no evidence of a Ce$_2$Co$_{17}$ phase within the sample. Instead, cell boundaries appear to host a modified CeCo$_5$ phase with ordered Cu substitution on Co-sites. We propose that this high-density network of Cu concentration gradients forms an array of effective nanoscale domain wall pinning sites. This aligns with previous findings that Cu significantly alters domain wall energy in Sm(Co$_{1-x}$Cu$_x$)$_5$ magnets [25]. Thus, with increasing Cu content, the coercivity mechanism in Ce(Co$_{1-x}$Cu$_x$)$_5$ shifts from a nucleation-dominated to a pinning-dominated regime at roughly 25 at.% Cu [3], driven by the emergence of a nanoscale, compositionally modulated nanostructure.

# 2 Methods

## 2.1 Synthesis and mesoscale characterization
High-purity elemental metals - cerium (Ce), cobalt (Co), and copper (Cu), each with a purity above 99% - were weighed according to the nominal composition Ce(Co$_{0.8}$Cu$_{0.2}$)$_{5.4}$ and subsequently alloyed by induction melting in argon atmosphere. The resulting 50g alloy ingot was sealed, under argon atmosphere, in quartz ampoules for heat treatment. Homogenization and aging temperatures were selected based on the pseudo-binary $Ce(Co_{1-x}Cu_x)_5$ (x = 0...1) phase diagram [22], targeting the fully miscible region for homogenization and the miscibility gap for aging. The 50 g ingot of $Ce(Co_{0.8}Cu_{0.2})_{5.4}$ was homogenized at 1100 °C for 2 hours, followed by rapid quenching in water. The ingot was then divided into two parts. The first part, used directly after the homogenization treatment, is referred to as the low coercivity (low $H_c$) sample. The second part underwent an aging treatment at 400°C for 4 hours, followed by water quenching, and is referred to as the high coercivity (high $H_c$) sample. $Ce(Co_{0.8}Cu_{0.2})_{5.4}$ nearly single crystalline grains of 1-2 mm in size were extracted from both, the homogenized and annealed ingots using the preparation route described in [8]. The term "nearly single crystalline" highlights that the samples contain two variations of the $Ce(Co_{1-x}Cu_x)_5$ (1:5) phase, which differ in their degree of Cu-ordering but are structurally coherent and closely related, as discussed below. The crystal quality of the grains was confirmed by Laue diffraction. Around 1 mm large nearly single crystalline" samples of $Ce(Co_{0.8}Cu_{0.2})_{5.4}$ were measured in a vibrating sample magnetometer (VSM) integrated in a physical property measurement system (Quantum Design, PPMS-14). For microstructure analyses, polycrystalline samples with a diameter of around 1cm were embedded in a conductive epoxy, mechanically grinded and polished. Surface morphology and crystallographic orientation at the micrometer scale were characterized using scanning electron microscopy (SEM, FEI Helios Nanolab 600i) in backscatter electron (BSE) and secondary electron (SE) modes, along with electron backscatter diffraction (EBSD, Zeiss SIGMA 500). Chemical composition was measured using energy dispersive X-ray spectroscopy (EDX, TESCAN VEGA3).

The crystal structures were determined using X-ray diffraction for powder samples (XRD, STOE diffractometer in transmission mode with a molybdenum K$_\alpha$ source) and backscattered Laue diffraction for nearly single crystalline samples.

## 2.2 Nanoscale composition and microstructure analysis

For the nanoscale composition analysis, needle-shaped specimens were extracted using a focused ion beam (FIB) from selected regions of the polished alloy surface in a dual-beam SEM/FIB system (FEI Helios Nanolab 600i). The procedure by Thompson *et al.* [26] was followed, with final milling performed with a low energy 5 keV Ga beam to minimize beam induced damage. These needle-shaped specimens were investigated using a CAMECA LEAP 5000 XR atom probe, equipped with a local electrode and a reflectron system. Analyses were performed at 40 K under ultra-high vacuum conditions ($10^{-10}$ mbar) using a pulsed UV laser (355 nm wavelength, 10 ps pulse duration, 40 pJ pulse energy, 200 kHz pulse rate, and 1-3% detection rate), yielding spatial and chemical data for ~50 million atoms per specimen. The analysis of APT data was performed with the AP Suite software v.6.3 by CAMECA.

A TEM lamella for conventional TEM imaging studies was prepared by cutting a slice of approximately 100 μm thickness of a bulk sample. Subsequent thinning down to an electron transparent lamella was achieved by a two-step ion milling process using a Gatan 691 Precision Ion Polishing System (PIPS). First the angles and ion beam energy were set to 8° and 5.5 keV, respectively, and milling was done until a small hole was observed in the sample center. In the second step, an angle of 2° and an ion beam energy of 2 keV were used in order to remove the damage from the previous step of milling. Bright-field (BF) TEM imaging and selected-area electron diffraction (SAED) measurements were carried out in a conventional transmission electron microscope (JEOL JEM 2100F). High-resolution high angle annular dark field (HAADF) scanning TEM (STEM) imaging was carried out in an aberration-corrected system (JEOL JEM-ARM200F) operated at 200 kV.

Electron-transparent specimens for magnetic imaging TEM studies were prepared using Ga+ sputtering and a conventional lift-out method with a dual-beam SEM/FIB system (FEI Helios Nanolab 600i). The TEM specimens were imaged at remanence under magnetic-field-free conditions (Lorentz mode) using a spherical aberration-corrected microscope operated at 300 kV (FEI Titan 80-300). Fresnel defocused images were acquired in Lorentz mode, where contrast at a defocus of $\delta z$ revealed bright (convergent) and dark (divergent) features at the locations of magnetic domain walls (DWs). Magnetic field was applied to the TEM specimen using the conventional objective lens of the microscope. Fresnel images and electron holograms were recorded using a direct-electron counting detector (Gatan K2-IS), and processed with the Gatan Microscopy Suite software. The electron biprism voltage typically ranged from 90 to 130 V, producing a fringe spacing of 3 nm with a contrast of 40%.

## 2.3 Micromagnetic simulations

Micromagnetic simulations were carried out using the open-source GPU-accelerated finite-difference (FD) program mumax3 [27]. We consider the free energy $F$ of the magnetic system with the domain volume $\Omega$ as

$$F(\vec{M}, \overrightarrow{H_{\text{ext}}}) = \int_\Omega \left[ \frac{A}{M_s^2}(\vec{\nabla}\cdot\vec{M})^2 - \frac{K_u}{M_s^2}(\vec{u}\cdot\vec{M})^2 - \mu_0\left(\frac{1}{2}\overrightarrow{H_{\text{dm}}}\cdot\vec{M} + \overrightarrow{H_{\text{ext}}}\cdot\vec{M}\right) \right] dV,$$

where $\vec{u}$ is the uniaxial vector and is identical to the magnetocrystalline easy axis for all coherent phases. $\vec{H}$ and $\vec{M}$ are respectively the magnetic field and magnetization vectors, $\overrightarrow{H_{\text{ext}}}$ is the external field, and $\overrightarrow{H_{\text{dm}}}$ is the demagnetizing field.

The 1st order uniaxial anisotropy constant $K_u$, and the saturation magnetization $M_s$ are concentration-dependent parameters and were measured for a series of Ce(Co$_{1-x}$Cu$_x$)$_{5.4}$ (x=0…0.3) single grain samples: The temperature-dependent measurements are shown in the supplementary **Figure S1**, and fitted room temperature values appear in **Figure 1b₁** and **b₂**.

For simplicity the exchange stiffness parameter $A$ was approximated as Cu-independent and calculated using the relation $l_{\text{dw}} = \pi\sqrt{A/K_u}$, where the 180° Bloch-type domain wall width of $l_{\text{dw}} = 4.3 \pm 0.6$ nm was determined by TEM magnetic imaging (cf. **Figure 6**f), and the uniaxial anisotropy constant of $K_u = 1.46$ MJm$^{-3}$ of Ce(Co$_{0.8}$Cu$_{0.2}$)$_{5.4}$ was taken (**Figure 1b$_1$**, $x = 0.2$). The magnetization evolution was computed by the over-damped Landau-Lifshitz equation, based on the steepest conjugate gradient (SCG) method [28,29] with:

$$\frac{\vec{m}^{(i+1)} - \vec{m}^{(i)}}{\Delta^{(i)}} = \vec{m}^{(i)} \times \frac{1}{\mu_0 M_s}\left[\vec{m}^{(i)} \times \frac{\delta \mathcal{F}}{\delta \vec{m}^{(i)}}\right],$$

$$\vec{m} \equiv \vec{M}/M_s, \text{ subject to } |\vec{m}| = 1,$$

where $\Delta^{(i)}$ is the iteration step size at $i$-th iteration.

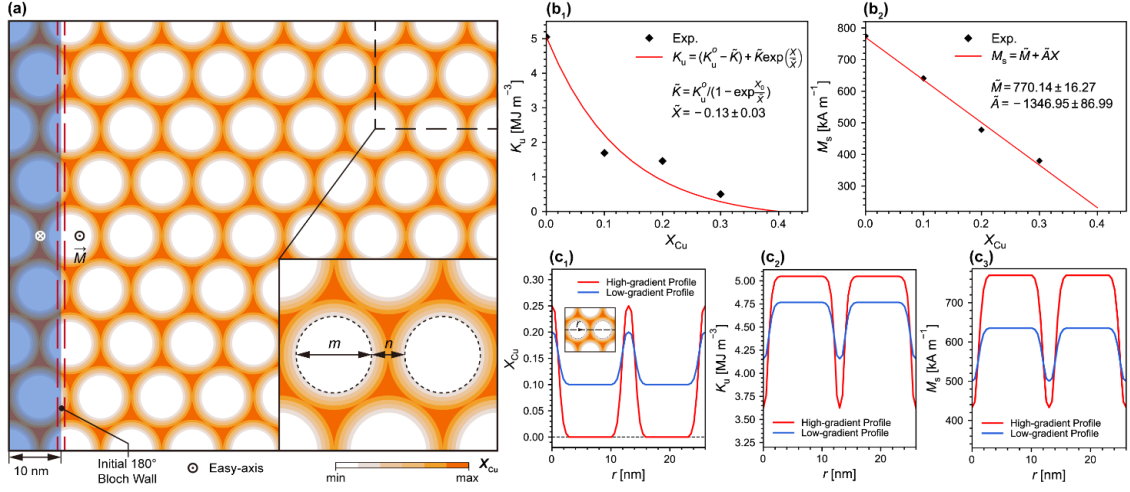

**Figure 1 Setup of the micromagnetic simulations:** (a) Schematic of the parameterized nanostructure with a 180° Bloch wall along $z$-axis (out-of-plane the easy axis). An external magnetic field is applied along $z$-direction to simulate the magnetization reversal. (b$_1$-b$_2$): experimentally fitted Cu concentration dependency of the 1$^{st}$ order uniaxial anisotropy constant $K_u$ and the saturation magnetization $M_s$, respectively. (c$_1$-c$_3$) Spatial profiles across the concentration-gradient regions of (c$_1$) Cu concentration $x = X_{Cu}$ in Ce(Co$_{1-x}$Cu$_x$)$_{5.4}$, (c$_2$) $K_u$ and (c$_3$) $M_s$.

The simulation domains with a size of $104 \times 92 \times 8$ nm³ were discretized by the finite difference (FD) meshes with a grid size of $0.5 \times 0.5 \times 0.5$ nm³. Based on nanostructures observed experimentally, Ce(Co$_{0.8}$Cu$_{0.2}$)$_{5.4}$ was modeled as a periodic structure comprising Cu-poor circular cells (diameter $m = 8$ nm), surrounded by the concentric concentration-gradient regions of $n/2$ in width (and $n = 5$ nm), embedded in the Cu-rich "matrix", representing cell boundaries (**Figure 1a**). Maximum and minimum Cu concentrations $X_{Cu}$ were approximated as constants based on frequent values observed by APT. The $X_{Cu}$-profiles in the concentration gradient regions were modelled using Gaussian functions (**Figure 1c$_1$**), and used to determine the spatially dependent $K_u$ and $M_s$ profiles, shown in **Figure 1c$_2$** and **1c$_3$**, based on fitted data $K_u(X_{Cu})$ and $M_s(X_{Cu})$ (**Figure 1a$_1$** and **a$_2$**).

Based on input from EBSD and APT, discussed below, the easy axis of the nanostructure was oriented along the $z$ (out-of-plane) direction. The initial configuration included a reversed magnetic domain with magnetization vector $\vec{M}$ is anti-parallel to the easy axis, and a with a 180° Bloch wall.

As the $\overrightarrow{H_{\mathrm{dm}}}$ is calculated directly by the magnetostatic convolution of magnetization configuration over the simulation domain [27,30], geometry of the simulation domain as well as the boundary condition can significantly affect the resolved demagnetization process. To emulate the demagnetization field in the bulk, the periodic boundary condition (BC) was applied on the two boundaries perpendicular to the $z$ direction by macro geometry approach [31], while the Neumann BC was applied on other boundaries [27]. It is worth noting that a simulation domain with gridsize $(N_x, N_y, N_z)$ with the non-zero PBC factors $(P_x, P_y, P_z)$ by macro geometry approach correspond approximately to a domain of gridsize $(2P_x N_x, 2P_y N_y, 2P_z N_z)$. In this case the only non-zero PBC factor $P_z = 4$, given the deviation of 5% in-plane demagnetizing factor from the analytical one on an infinitely long domain [27].

## 3 Results

### 3.1 Mesoscale microstructure characterization

**Figure 2a** shows the magnetic hysteresis curves, including the initial magnetization curves, for the homogenized and aged samples. The homogenized low $H_c$ sample exhibits a coercivity of $H_c$ = 0.48 T and a remanence of $M = 50$ Am²/kg. The aged high $H_c$ sample shows $H_c$ = 1.02 T with a slightly lower $M = 41$ Am²/kg. Reported values are independent of sample shape, as hysteresis data were corrected for the demagnetization factor. In both cases, the coercivity mechanism is pinning-type, as evidenced by the flat slope of the initial magnetization curves at low fields and the partial hysteresis loops shown in supplementary **Figure S2a-b**. A closer examination of the initial magnetization curves reveals a two-step change in slope for the high $H_c$ sample, with a depinning field of 1 T. In contrast, the low $H_c$ sample exhibits a single-step slope change at a lower depinning field of 0.4 T, consistent with the observed coercivity trend. Additional magnetometry analyses including low temperature loops and $B$-$H$ loops can be found in supplementary **Figure S2**.

XRD and Laue diffraction results for a single grain of the high $H_c$ sample are shown in **Figure 2b** and **2c**. The Miller indices of the Ce(Co$_{0.8}$Cu$_{0.2}$)$_{5.4}$ peaks with high intensity are indicated and the Bragg peak positions for CeCu$_5$ as well as a simulated CeCo$_5$ pattern are provided for comparison. The peaks of Ce(Co$_{0.8}$Cu$_{0.2}$)$_{5.4}$ align with CeCo$_5$, showing a minor left-shift likely due to Cu-substitution, as the unit cell of CeCu$_5$ is ~11% larger than CeCo$_5$ [15,16], and a slight Ce deficiency. The Laue diffraction pattern of Ce(Co$_{0.8}$Cu$_{0.2}$)$_{5.4}$ in **Figure 2c** shows no evidence of secondary phase or twinning, confirming that the aged, off-stoichiometric sample retains the CeCo$_5$ structure.

**Figure 2d** and **2e** show representative BSE SEM images acquired from single grains of the high $H_c$ and low $H_c$ samples, respectively. Both images confirm homogeneous, single-phase state at the mesoscale. The composition Ce(Co$_{0.81}$Cu$_{0.19}$)$_{5.33}$, determined by EDX, closely matches the nominal stoichiometry. BSE SEM and EBSD analyses of the polycrystalline high $H_c$ sample (supplementary **Figure S3**) reveal large grains of ~100-500µm with a high degree of texture and low misorientation angles < 3°.

Since no mesoscale differences in microstructure or composition are observed between the samples, yet they exhibit different coercivities, we expect differences to arise at the nanoscale in terms of composition and/or structure, similar to the precipitates reported by Leamy *et al.* [21] for Ce(Co,Fe,Cu)$_5$. To investigate this, we employ a combination of high-resolution structural, chemical and magnetic analyses using TEM and APT.

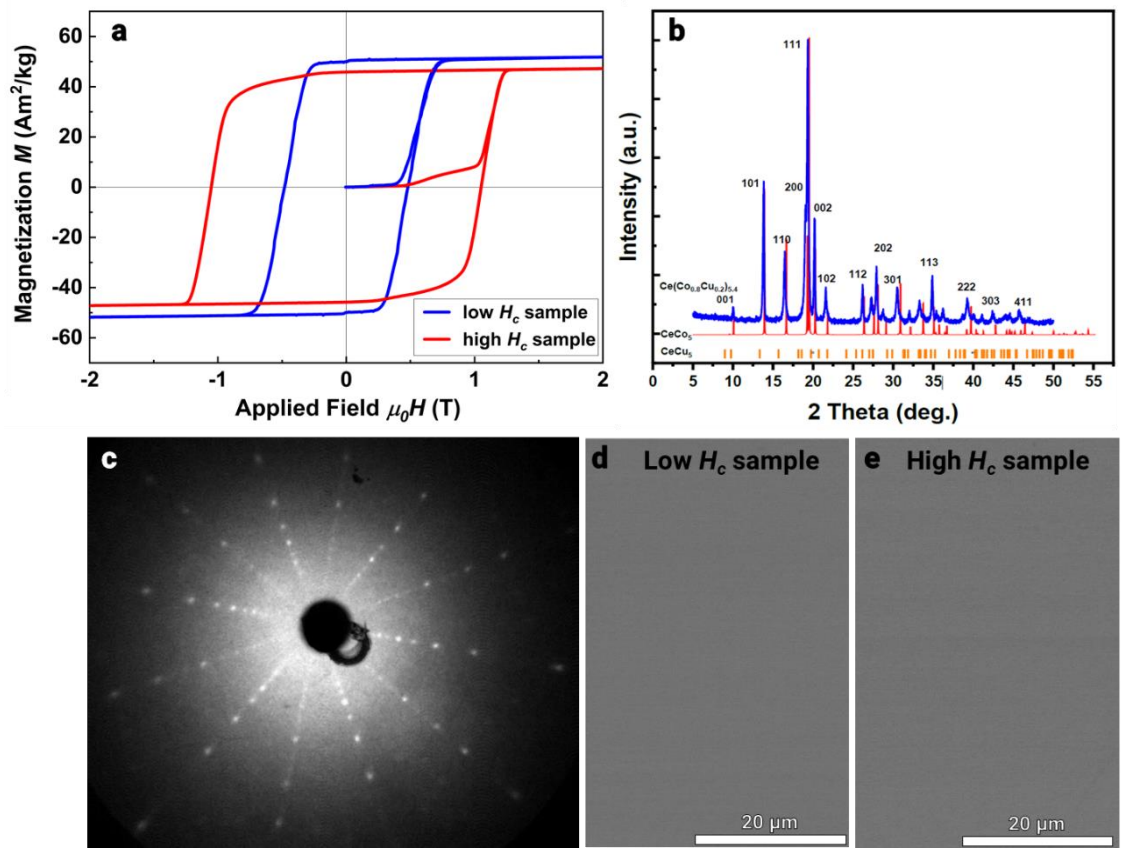

**Figure 2 Structural and magnetic characterization of Ce(Co$_{0.8}$Cu$_{0.2}$)$_{5.4}$ single grains after different heat treatments: homogenized low $H_c$ sample and aged high $H_c$ sample**. (**a**) Hysteresis and initial magnetization curves for the low and high $H_c$ samples. (**b**) XRD patterns for the high $H_c$ sample (blue) with a theoretical CeCo$_5$ pattern (red) and CeCu$_5$ peak positions (orange). (**c**) Laue diffraction pattern for the high $H_c$ sample. (**d-e**) BSE-SEM images for low and high $H_c$ samples, respectively.

## 3.2 High-resolution compositional mapping

### 3.2.1 Comparison of nanostructure and composition in low $H_c$ and high $H_c$ samples

**Figure 3** shows the APT analysis of the low and high $H_c$ samples. Their compositions, Ce(Co$_{0.82}$Cu$_{0.18}$)$_{5.27}$ and Ce(Co$_{0.83}$Cu$_{0.17}$)$_{5.18}$, match the nominal stoichiometry with a slight Ce excess (cf. supplementary **Table S1**), consistent with EDX. While 3D reconstructions, shown in **Figure 3a** and **3b**, appear similar, 2D Cu concentration maps, extracted from 10 nm thick slices through the reconstructed 3D point cloud, reveal significant differences (**Figure 3c** and **3d**): both samples show quasi-periodic Cu fluctuations consisting of Cu-poor and Cu-rich regions, but the high $H_c$ sample spans a wider range of $0-35$ at.% Cu (violet-yellow contrast in **Figure 3d**), whereas the low $H_c$ sample only varies between $10-20$ at.% Cu (blue-green contrast **Figure 3c**), indicating stronger local inhomogeneity for the high $H_c$ sample. This is also reflected in the average compositions of these regions (cf. supplementary **Table S1**): in the low $H_c$ sample, the Cu-poor regions are ~4 at.% Cu below average and Cu-rich regions ~6 at.% above average, corresponding to Ce(Co$_{0.86}$Cu$_{0.14}$)$_{5.10}$ and Ce(Co$_{0.75}$Cu$_{0.25}$)$_{5.42}$, respectively. In contrast, the fluctuations are more pronounced in the high $H_c$ sample, with Cu-poor regions being ~6 at.% below and Cu-rich regions ~6 at.% Cu above average, translating to Ce(Co$_{0.90}$Cu$_{0.10}$)$_{4.95}$ and Ce(Co$_{0.76}$Cu$_{0.24}$)$_{5.34}$.

This is further confirmed by 1D composition profiles (**Figure 3e** and **3f**) that show a higher maximum Cu concentration of ~40 at.% in the high $H_c$ sample compared to ~30 at.% in the low $H_c$ sample, corresponding to Ce(Co$_{0.56}$Cu$_{0.44}$)$_{5.25}$ and Ce(Co$_{0.70}$Cu$_{0.30}$)$_{5.22}$. For both samples, the local Ce concentration varies only slightly ($16 \pm 5$ at.%), whereas Co and Cu concentrations show high variance ($\pm 20$ at.%) and anti-correlation, visually seen as "mirrored" Co/Cu graphs. These findings are consistent with CeCo$_5$-type ordering, where Cu occupies Co-sites on the transition metal sublattice, while the Ce rare-earth sublattice remains mostly undisturbed, supporting the XRD and Laue results.

Relative concentration frequency analysis, which considers the full volume of APT data [32], quantitatively confirms stronger Cu-Co separation in the high $H_c$ sample, as shown in **Figure 3g**. Cu-poor (0–10 at.%) and Cu-rich (25–40 at.%) regions are more prevalent, while intermediate Cu concentrations (10–25 at.%) are underrepresented, as reflected in the distributions in the top graph and more clearly highlighted by the differences in the bottom graph. Co shows a corresponding mirrored trend, whereas Ce distributions remain nearly identical between the samples.

Morphologically, the low $H_c$ sample contains irregular features on the length scale of 5 nm without any apparent spatial pattern (**Figure 3c**). In contrast, the high $H_c$ sample displays larger, aligned features with a stronger chemical separation (**Figure 3d**): elongated Cu-poor regions, 5-10 nm thick and ~20 nm long, exhibit a preferential alignment with their long axes nearly parallel to one another, separated by ~5 nm thick Cu-rich regions. Note that the APT analyses presented so far were per-formed on specimens prepared from grains of a non-specifically selected crystallographic orientation. The *c*-axis direction could not be determined retrospectively through APT-based crystallographic analysis [33].

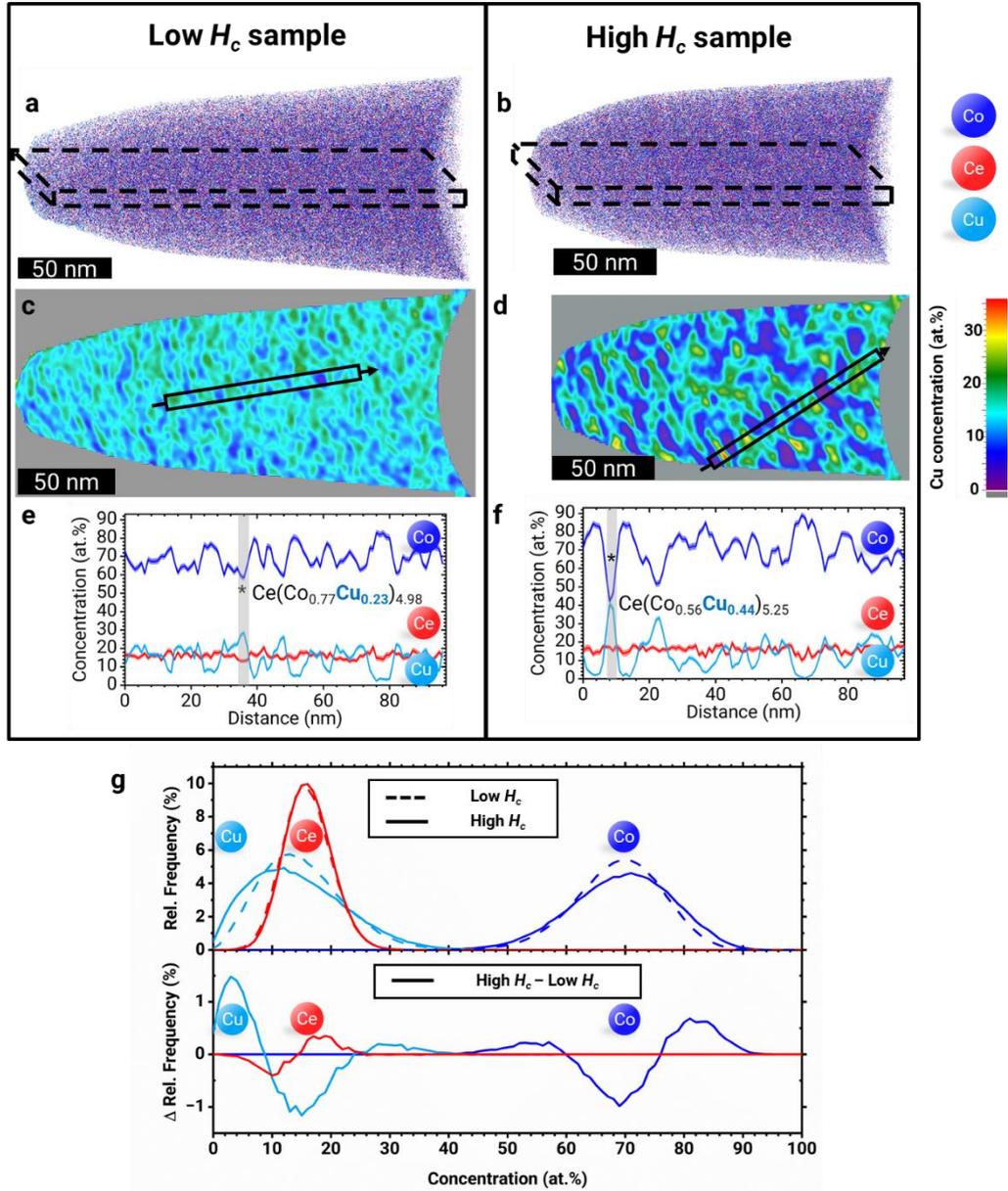

**Figure 3 3D APT data comparing low $H_c$ and high $H_c$ samples of Ce(Co$_{0.8}$Cu$_{0.2}$)$_{5.4}$. (a-b)** 3D APT reconstructions of low $H_c$ and high $H_c$ samples, respectively, with Co(dark blue), Ce (red), Cu(light blue). **(c-d)** 2D Cu concentration maps from 10nm thick slices indicated in (a-b) revealing chemical modulation. **(e-f)** 1D concentration profiles extracted from cylinders (5 nm diameter, 100 nm length) highlighted in (c-d). **(g)** Relative concentration frequency analysis for each element, calculated from 100-atom cubic volumes, showing compositional distributions $f$ (top), and distribution differences $\Delta f = f_{\text{high } H_c} - f_{\text{low } H_c}$, (bottom).

### 3.2.2 Crystallographic orientation analysis of the nanostructure in the high $H_c$ sample

To determine the crystallographic alignment of the elongated Cu-rich and Cu-poor features, EBSD was performed first in order to prepare APT specimens with the [001] direction of the CeCo$_5$ structure nearly parallel to the specimen axis, as demonstrated in **Figure S4**. APT crystallographic analysis [33] confirms this orientation, as described in **Figure S5**.

**Figure 4** compiles APT data from the crystallographically oriented high $H_c$ sample, with the *c*-axis indicated. In 3D reconstructions in **Figure 4a** and **4b**, elongated Cu-poor and Cu-rich features align closely with the *c*-axis, with deviation angles below 20°. Three ~5 nm thick slices were extracted from the reconstructions for further analysis.

The xy slice in **Figure 4c-d** shows a cross-section view of the structure, consisting of distorted oval Cu-poor *cells*, $5 \times 10$ nm$^2$ in size, containing around 0 at.% Cu, separated by ~5nm thin Cu-rich *cell boundaries*, exhibiting ~20-30 at.% Cu. Local Cu maxima exceeding 30 at.% occur at intersections of three cell boundaries, as shown in the 2D Cu concentration map in **Figure 4d**. These points potentially act as stronger pinning sites for magnetic domain walls, as the coercivity of the isostructural SmCo$_5$ phase is known to increase as a function of Cu concentration [11]. Rare-earth-to-transition-metal ratios, Ce/(Co+Cu), derived from 1D profiles (**Figure S6**) fluctuate around $1:5 = 0.2$, indicating the predominance of the 1:5 phase in the sample. Only in isolated, Cu- and Ce-deficient cells (cf. 2D chemical maps in the supplementary **Figure S7**), does this ratio approach $2/17 \approx 0.12$, suggesting occasional local 2:17-like cells. The xz slice in **Figure 4e-f** shows the side view of the structure, with roughly rectangular, ~20nm long cells, which can be approximated as Cu-poor cylinders, separated by Cu-rich boundaries. The yz slice in **Figure 4g-h**, taken near a "pole" of APT data with enhanced spatial resolution [34], reveals Co-rich planes that are continuous across the cell and cell boundary interface, indicating crystallographic coherency. Note that the Co-Co distance, determined from the spatial distribution map [35] (inset of **Figure 4h**), was calibrated to match the *c* lattice constant of CeCo$_5$ ($c = 0.4019$ nm [1]), identifying the Co-rich planes as (001) planes.

The stronger chemical segregation in the high $H_c$ sample indicates a connection between pinning strength and the degree of chemical segregation of the cellular structure. TEM microstructure and magnetic imaging were performed on the high $H_c$ sample to further investigate this mechanism.

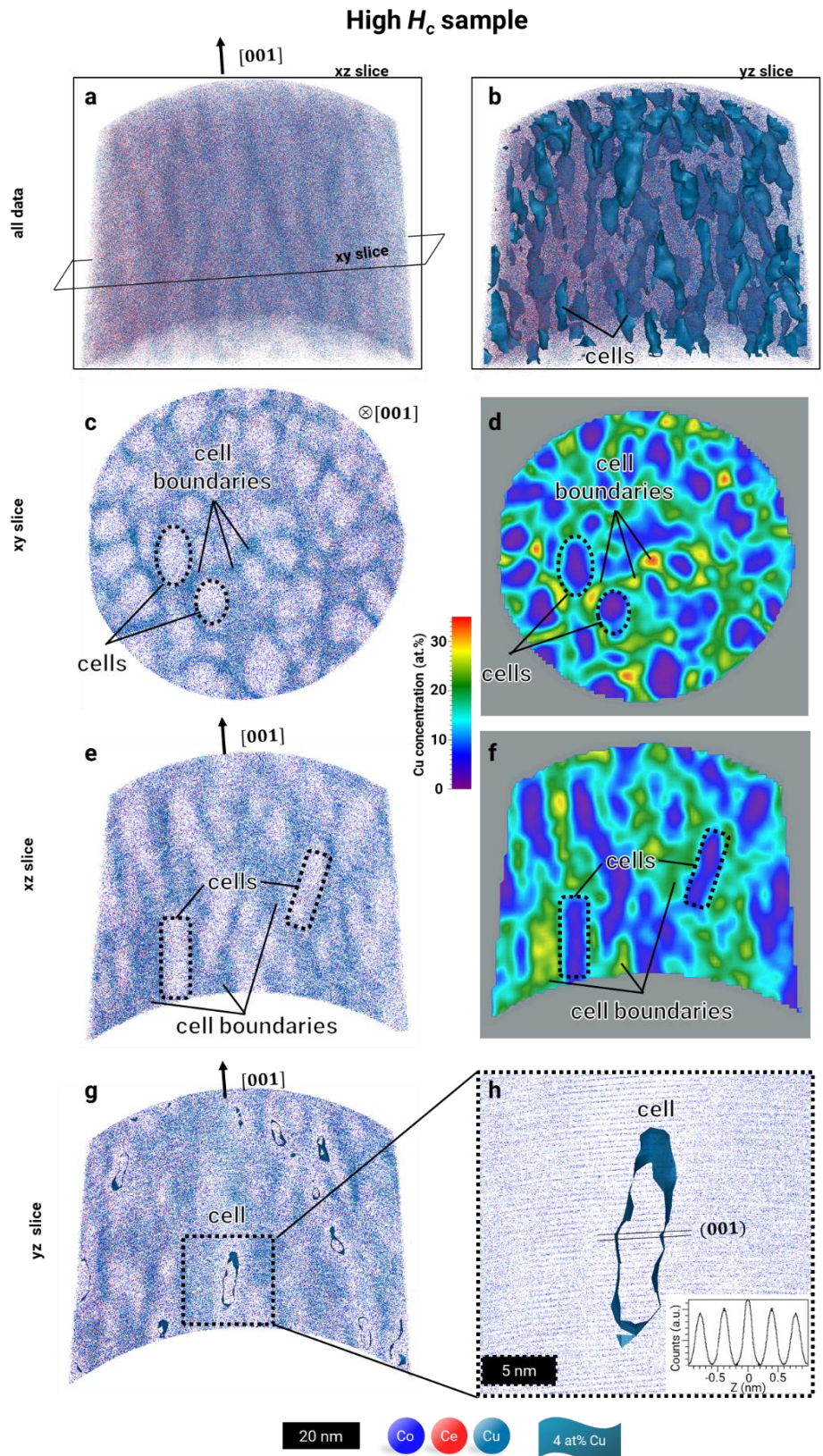

**Figure 4 APT analysis of the crystallographically oriented high $H_c$ Ce(Co$_{0.8}$Cu$_{0.2}$)$_{5.4}$ sample**, with the [001] axis oriented almost parallel to *z*-axis of the APT reconstruction. **(a-b)** 3D APT reconstructions without and with visualizing Cu-poor cylindrical features (cells) using a 4 at.% Cu isovalue, respectively, with atoms of Co (dark blue), Ce (red), Cu (light blue). **(c-d)** 3D reconstruction and 2D Cu concentration map of the xy slice (10 nm thick, extracted from (a) with c-axis out of plane) revealing a cross-sectional view on the cellular nanostructure with Cu-poor cell and Cu-rich cell boundaries.

**(e-f)** 3D reconstruction and 2D Cu concentration map of the xz slice (10 nm thick, extracted from (a) with c-axis in plane), showing elongated cylindrical morphology with preferential orientation to the c-axis in side view. **(g-h)** Low and high magnification 3D reconstruction of the yz slice, extracted from (b) with c-axis in-plane, close to the (001) pole, revealing Co-rich (001) atomic planes that are continuous across the cell/cell boundary interface. Inset: Spatial distribution map from a 3x5x35 nm$^3$ cube used to extract Co-Co spacing.

## 3.3 High-resolution microstructural and magnetic imaging of the high $H_c$ sample

### 3.3.1 Microstructure imaging

**Figure 5** presents a comprehensive TEM analysis of the high $H_c$ Ce(Co$_{0.8}$Cu$_{0.2}$)$_{5.4}$ sample, imaged with c-axis out of plane. In **Figure 5a**, a bright-field TEM image shows a chemically homogeneous matrix. Closer inspection of the higher magnification BF-TEM image from the same area along the [001] zone axis, shown in **Figure 5b**, reveals contrast variations on the ~10 nm scale. The observed contrast could be due to compositional variation, consistent with the chemically modulated cellular structure observed via APT. The atomic-resolution HAADF STEM image (**Figure 5c**) matches the hexagonal CeCo$_5$ (P6/mmm) structure. The corresponding fast Fourier transformed (FFT) image pattern (**Figure 5d**) agrees with the simulated pattern (**Figure 5g**), showing no additional reflections. This suggests that Cu atoms are randomly distributed on the Co-sublattice in the CeCo$_5$ structure (**Figure 5f**), which is reasonable given the small atomic radius difference between Co and Cu of less than 3%.

However, a high-index diffraction pattern acquired along the [-1-12] zone axis (**Figure 5e**) reveals additional satellite reflections, indicating the presence of a structurally modified phase in the matrix of the sample. Similar satellite peaks were consistently observed in other high-index diffraction patterns of the matrix (see supplementary **Figure S8**). Although no structural model was able to reproduce these satellite peaks observed for the [-1-12] zone axis or higher zone axes, insights into their origin were gained by analyzing the Cu-rich GB precipitates in the same sample. Diffraction patterns from these precipitates, acquired along the [001] zone axis, also contained similar satellite reflections, as shown in supplementary **Figure S9**. In this case, the satellite peaks were successfully reproduced in a simulated diffraction pattern (**Figure 5i**) by assuming a structurally modified CeCo$_5$ (P6/mmm) phase, referred to as Cu-ordered CeCo$_5$ phase. This phase features an ordered arrangement of Cu atoms on Co sites on one of the six symmetry-equivalent [100] directions, as depicted in **Figure 5h**. Although the satellite peaks in the [001] diffraction pattern could also be interpreted as originating from the Ce$_2$Co$_{17}$ phase as proposed by Leamy et al. [21], our data supports the CeCo$_5$-type structure for three reasons: (1) the Ce distribution observed in STEM-EDX is uniform (supplementary **Figure S9**); (2) the rare-earth-transition-metal ratio determined by APT is $\frac{Ce}{Co+Cu} \sim \frac{1}{5.4}$, consistent with the intended stoichiometry (supplementary **Figure S6**); and (3) the satellite peaks observed in higher-order diffractions patterns (supplementary **Figure S8**) are incompatible with the presence of the Ce$_2$Co$_{17}$ phase.

Since satellite peaks appeared in the diffraction patterns of both, the matrix and GB precipitates, we suggest that the Cu-ordered CeCo$_5$-phase is also present within the matrix. Given that Cu-ordering was associated with higher Cu-content in GB precipitates, we propose that this Cu-ordered CeCo$_5$ phase is localized in the Cu-rich cell boundaries of the matrix, as observed by APT. In contrast, Cu-poor cells likely exhibit a conventional CeCo$_5$ phase. Therefore, diffraction patterns from the matrix likely represent a superposition of the two structural variants of the CeCo$_5$ phase. Interestingly, the ordering-related satellite peaks are not visible [001] zone axis, but appear only in higher order zone axis – a phenomenon that requires further analyses.

Note that the cellular structure in the matrix on the nm scale, as observed by APT, could not be directly resolved by TEM due to a relatively thick lamella of about 100 nm, limiting resolution of the fine chemical modulation.

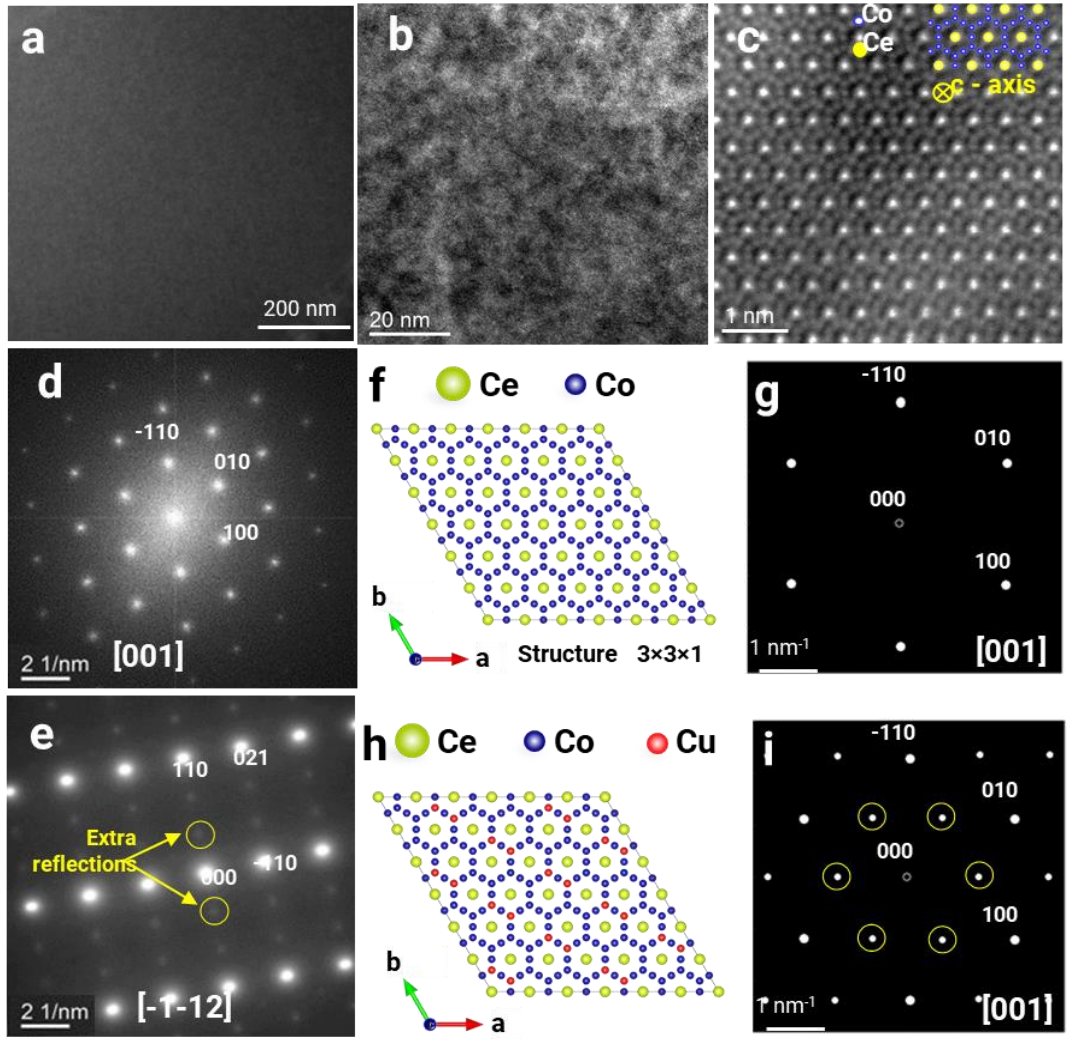

**Figure 5 TEM characterization of the Ce(Co$_{0.8}$Cu$_{0.2}$)$_{5.4}$ high $H_c$ sample: (a)** Bright-field TEM image showing a homogeneous matrix. **(b)** High-magnification BF-TEM image and **(c)** atomic-resolution HAADF STEM images of the matrix (detector semi-angle 90–370 mrad). **(d)** Fast Fourier transformed image of **b**, **(e)** High-index diffraction pattern along the [-1-1 2] zone axis showing satellite reflections, as indicated by yellow arrows. **(f,g)** Structural models with 3×3×1 cells of simple CeCo$_5$ and CeCo$_5$ with Cu-ordering in lines on Co-cites. **(g,i)** Simulated diffraction along the [001] zone axis using the structural models in (f,h).

### 3.3.2 Magnetic domain imaging

**Figure 6** shows the TEM-based magnetic domain imaging in the Ce(Co$_{0.8}$Cu$_{0.2}$)$_{5.4}$ high $H_c$ sample, analyzed with the c-axis in-plane. In **Figure 6a**, the HAADF STEM image reveals compositional contrast associated with Cu and Co distribution. The corresponding fast Fourier transform (FFT) pattern (inset of **Figure 6a**), confirms the [0-10] zone axis of the hexagonal (P6/mmm) structure. The atomic-resolution HAADF STEM image (**Figure 6b**) was processed using a high-pass filter to enhance compositional contrast: Cu-rich regions ($Z_{Cu} = 29$) appear blue, and Co-rich regions ($Z_{Co} = 27$) appear red, according to HAADF STEM intensity scaling with the atomic number as $I \propto Z^2$. These observations are consistent with APT results (**Figure 4d1 and d2**) showing Cu-rich cell boundaries and Cu-poor cells.

Mesoscale domain structure imaging by MFM with the nominal c-axis out of plane for the high $H_c$ sample, shown in Supplementary **Figure S10**, reveals fine magnetic domains approximately 0.5 μm in size. This fine domain structure is more characteristic of precipitation-hardened 2:17-type SmCo magnets with a cellular structure [36] than of single-crystalline SmCo$_5$ magnets [3,23], consistent with the cellular structure observed by APT. The nanoscale domain structure is characterized using an underfocused Fresnel image, shown in **Figure 6c**, acquired under zero field conditions following a 1 T field application along the *c*-axis. The resulting magnetic domains are separated by characteristic zigzag domain walls (DW) that appear with bright/dark contrast in the Fresnel image. The zigzag-like pattern may arise from interactions between the DWs with the cellular nanostructures, whose long axes deviate up to 20° from the the *c*-axis, as observed by APT in **Figure 4**.

Off-axis EH [37] was further used to reconstruct the projected magnetic induction map using the electron wave phase shift $\phi$ (**Figure 6d**), in the DW region highlighted in **Figure 6c**. The magnetic and electrostatic contributions to the phase shift were not separated in this analysis, as the lamella thickness was uniform, resulting in a constant electrostatic contribution to the phase shift. The magnetic induction map reveals the field distribution around the DWs with segments of 180° and 90° magnetization rotation.

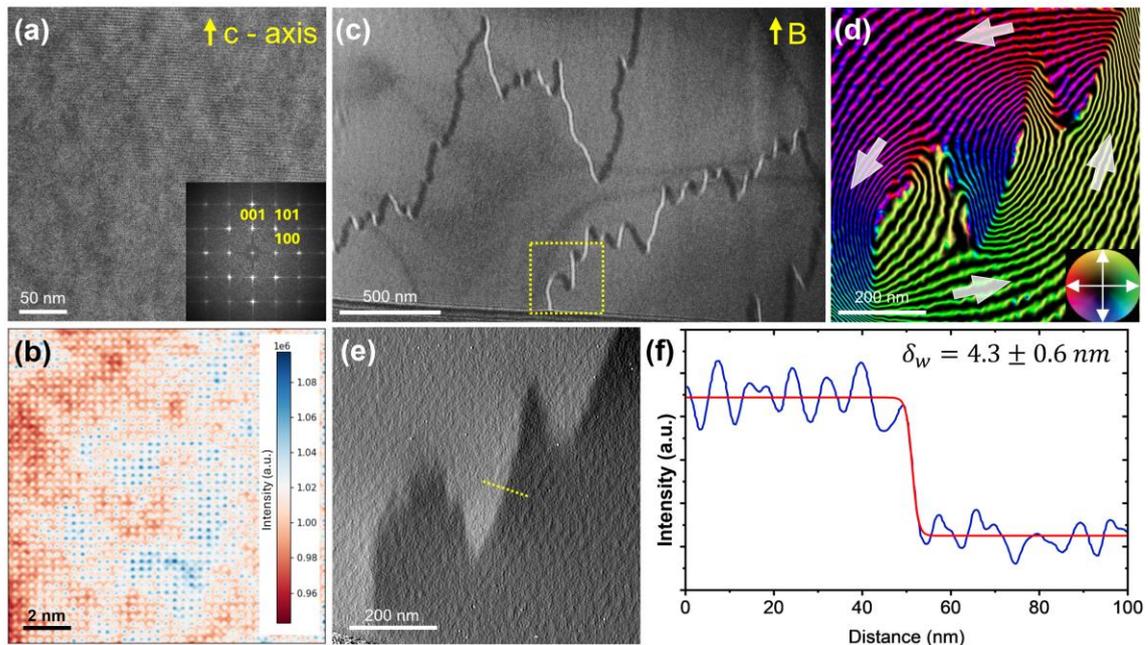

**Figure 6 Magnetic imaging of Ce(Co$_{0.8}$Cu$_{0.2}$)$_{5.4}$ high $H_c$ sample: (a)** HAADF STEM image and inset FFT pattern along [0-10] zone axis, showing chemical contrast. **(b)** High-pass-filtered atomic-resolution HAADF STEM image highlighting Cu-rich (blue) and Cu-poor (red) regions. **(c)** Fresnel Lorentz TEM image, showing black and white contrast at the positions of zigzag DWs, nucleated under applied

field B = 1T. The image is recorded in magnetic-field-free conditions at a defocus of 1 mm. **(d)** Magnetic induction map extracted from off-axis EH of a DW region from (c) indicated by a yellow dotted square. The contour spacing is π/5 rad. Arrows and colors indicate projected in-plane magnetic field directions. **(e)** Differential phase shift across the 180° DW used for **(f)** DW width measurement, with $\delta_w$= 4.3 ± 0.6 nm extracted by tanh-fitting.

To quantify the width of the 180° DW, the differential ($\partial\phi/\partial x$) of the recorded phase shift (**Figure 6e**) was fitted in the corresponding region using the hyperbolic tangent model of the form: $y = y_0 \pm a \cdot \tanh((x - x_0)/w)$, where $y_0$, $a$, $x_0$, and $w$ are constants obtained from the fit and the domain wall width $\delta_w$ is given by $\delta_w = \pi w$. The resulting DW width was determined to be 4.3 ± 0.6 nm (**Figure 6d**) for Ce(Co$_{0.8}$Cu$_{0.2}$)$_{5.4}$ high $H_c$ sample, comparable to reported values for NdFeB, 1:5-type SmCo and 2:17-type SmCo permanent magnets (2-6 nm) [38].

## 3.4 Micromagnetic simulation of high and low $H_c$ samples

Magnetometry, TEM and APT measurements indicated that the Cu/Co gradients in the cellular structure of Ce(Co$_{0.8}$Cu$_{0.2}$)$_{5.4}$ are correlated with more effective DW pinning, and consequently, higher coercivity. To understand the influence of Cu concentration $X_{\text{Cu}}$ and its spatial gradient on the magnetization reversal and the coercivity field $H_c$, we performed micromagnetic simulations on the parameterized nanostructures, based on our APT, TEM and magnetic imaging results, presented in Figure 7.

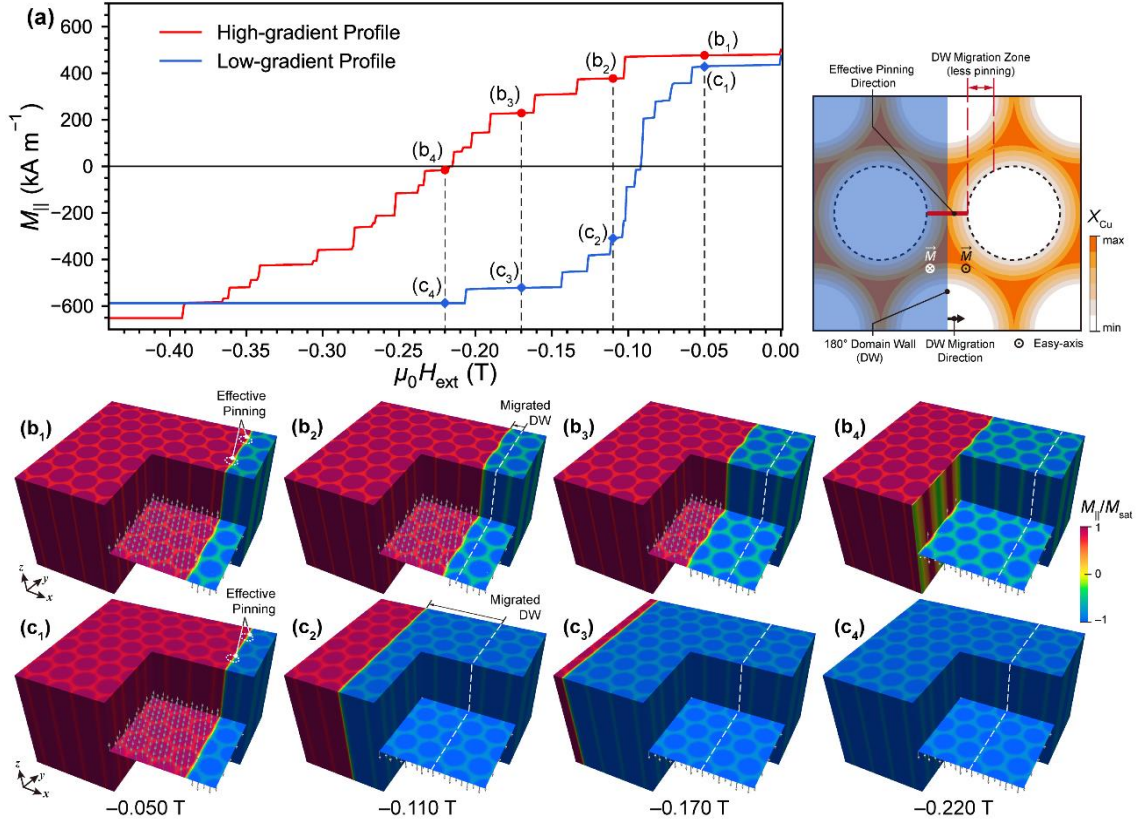

**Figure 7 Micromagnetic simulations comparing demagnetization in high and low $H_c$ Ce(Co$_{0.8}$Cu$_{0.2}$)$_{5.4}$ nanostructures, denoted as high-gradient and low-gradient $X_{Cu}$-profiles, respectively.** (**a**) Simulated demagnetization curves (left) and schematic of the nanostructure with the Cu-gradients, with highlighted effective pinning direction and domain wall (DW) migration zone (right). (**b$_1$-b$_4$**) High-gradient $X_{Cu}$ profile domain configurations at selected applied fields $\mu_0 H_{ext}$ marked in (a), with marked DW location and pinning sites. (**c$_1$-c$_4$**) Low-gradient $X_{Cu}$ profile domain configurations under the same conditions.

The parameterized nanostructure consists of the periodic Cu-poor cylindrical cells surrounded by Cu-rich cell boundaries, we the Cu-concentration follows a continuous gradient. Tilt angles of the cellular structure were neglected, and the dimensions of the cellular structure were kept constant. The spatially varying $X_{Cu}$ profiles were approximated assuming a constant minimum concentration $X_{Cu}^{min}$ in the cells and Gaussians peaking at a maximum concentration $X_{Cu}^{max}$ in the cell boundaries, with $X_{Cu}^{min}, X_{Cu}^{max}$ values derived from APT 1D profiles (see **Figure 3a3** and **b3**). Two representative cases were implemented:

(1) High-gradient profile: approximates the high-$H_c$ sample nanostructure with $X_{Cu}^{min} = 0.00$ and $X_{Cu}^{max} = 0.25$
(2) Low-gradient profile: mimics the low-$H_c$ sample $X_{Cu}^{max} = 0.20$ and $X_{Cu}^{min} = 0.10$

Micromagnetic parameters - magnetocrystalline anisotropy $K_u$ and the saturation magnetization $M_s$ – were modeled as functions of local Cu concentration $X_{Cu}$, in line with experimental results introduced in the *Methods* **section 2.3**. Since both parameters decrease with increasing Cu-content $X_{Cu}$, the gradient in $X_{Cu}$ leads to a spatial gradient in DW energy, according to $\sigma_{dw}^B \sim \sqrt{AK_u}$ for Bloch walls and $\sigma_{dw}^N \sim \sqrt{AM_s^2}$ for Néel walls, where $A$ is the exchange stiffness. The DW energy is expected to create local energy valleys at Cu-rich cell boundaries, acting as pinning sites during DW motion.

Figure 7$b_1$-$b_4$ and 8$c_1$-$c_4$ display the transient domain configurations of the high- and low-gradient profiles, respectively, at applied fields indicated in the demagnetization curves in Figure 7a. Both systems were initialized with an identical reversed domain configuration (cf. *Methods* **section 2.3, Figure 1a**), allowing a direct comparison of domain evolution. To simplify the simulation, a straight domain wall was assumed, disregarding the zigzag geometry observed via holography
At near-zero field of -0.05 T, the first jump of the curves represents the energetically preferred domain configurations (Figure 7$b_1$ and $c_1$) where the domain walls are effectively pinned at the Cu-rich boundaries. The lower remanent magnetization of the low-gradient structure resulting in an overall lower magnetization value. As the external field increases, multiple discrete magnetization jumps are observed in the demagnetization curves (Figure 7a), each corresponding to a DW depinning events from Cu-rich boundaries. The direction of DW propagation relative to the $X_{Cu}$ gradient is critical: DWs moving perpendicular to the gradient across the cell boundaries encounter stronger pinning in contrast to nearly free propagation within the Cu-poor cells denoted as DW migration zone. This alternation is visible in both structures, but depinning events are more frequent in the low-gradient structure.
At a higher field of -0.11 T (Figure 7$b_2$ and 8$c_2$), the DW in the low-gradient structure experienced seven depinning events and propagated significantly. In contrast, the DW of the high-gradient profile underwent only a single depinning event and hardly moved. The same trend is observed for higher fields (Figure 7$b_3$-$b_4$ and 8$c_3$-$c_4$) up to -0.22 T, implying less effective pinning and in the low-gradient structure. The coercivity of the high-gradient structure is roughly doubled ($\mu_0 H_c$ ~0.22T) compared to the low-gradient structure ($\mu_0 H_c$ ~0.09T), corroborating the experimental results. The underestimation of the simulated coercivity values likely arises from assumptions such as the presence of pre-nucleated reverse domains, the omission of structural effects due to Cu ordering, and uncertainties in the assumed intrinsic properties, as discussed in the following section in more detail.

Micromagnetic simulations confirm that a steeper Cu-gradient increases the pinning strength, and thereby coercivity, due to a higher DW energy contrast between Cu-poor cells and Cu-rich cell boundaries.

# 4 Discussion

In order to address the uncertainties surrounding the nanostructure and magnetization reversal mechanism in Ce(Co$_{0.8}$Cu$_{0.2}$)$_{5.4}$ reported in the literature, we compared two samples that had been subjected to different thermal treatments: a homogenized low $H_c$ sample with $\mu_0 H_c = 0.5$ T, and a homogenized and subsequently aged high $H_c$ sample with $\mu_0 H_c = 1$ T. This comparison raised four fundamental questions in this study:

(1) What is the nanostructure in Ce(Co$_{0.8}$Cu$_{0.2}$)$_{5.4}$ ?
(2) What is the driving force behind elemental demixing in the nanostructure?
(3) What is the coercivity mechanism in Ce(Co$_{0.8}$Cu$_{0.2}$)$_{5.4}$ ?
(4) How do the structure and coercivity mechanism compare to related systems, especially pinning-type Sm$_2$Co$_{17}$ magnets and magnets with giant intrinsic magnetic hardness?

*(1) Nanostructure*: At the mesoscale, both low $H_c$ and high $H_c$ Ce(Co$_{0.8}$Cu$_{0.2}$)$_{5.4}$ samples appear structurally and compositionally homogeneous, including the magnetic domain structure, as confirmed by BSE SEM, XRD, Laue diffraction and MFM. However, APT and TEM reveal a chemically and structurally modulated nanoscale cellular network in Ce(Co$_{0.8}$Cu$_{0.2}$)$_{5.4}$, as shown schematically in **Figure 8**. The high $H_c$ sample (**Figure 8a**) exhibits nearly cylindrical, Cu-poor cells, tilted up to 20° from the *c*-axis, enclosed by Cu-rich cell boundaries with coherent interfaces. The chemical contrast arises from the substitution of Co with Cu in the CeCo$_5$ lattice, in line with the formula Ce(Co$_{1-x}$Cu$_x$)$_{5.4}$. Sharp compositional gradients up to 12 at.% Cu/nm appear in the high $H_c$ sample. By contrast, the low $H_c$ sample (**Figure 8b**) shows weaker gradients of up to 8 at.% Cu/nm and lacks any apparent spatial pattern.

In the 1970s, Leamy *et al.* [21] attributed the presence of nanoscale precipitates in an Fe-doped Ce(Co$_{3.8}$Fe$_{0.5}$Cu$_{0.9}$) sample to the Ce$_2$Co$_{17}$ phase, based on additional TEM diffraction reflections absent in CeCo$_5$, although no compositional data were reported. While we observed similar diffraction patterns in both, the matrix and the ~100 nm large Cu-rich GB precipitates, our composition data are incompatible with the presence of the 2:17 phase in Ce(Co$_{0.8}$Cu$_{0.2}$)$_{5.4}$. STEM EDX revealed a uniform Ce-content throughout the sample, and APT derived Ce/(Co+Cu) fluctuated around $1/5 = 0.2$, clearly indicating the predominance of the 1:5 phase in the sample. Thus, the interpretation by Leamy *et al.* [21] is unlikely in our case. Instead, we propose that the additional reflections stem from a structurally modified Ce(Co,Cu)$_5$ phase, where Cu substitutes Co in an ordered manner, present in the matrix within the Cu-rich cell boundaries, as well as in GB precipitates, consistent with the TEM diffraction and APT data. This implies a structural, not just chemical, distinction between cells and boundaries, warranting further high-resolution TEM and 3D studies.

*(2) Nanostructure formation*: The thermodynamic driving force for chemical segregation in Ce(Co$_{0.8}$Cu$_{0.2}$)$_{5.4}$ is Gibbs free energy minimization, where compositional fluctuations are energetically more favorable than a homogeneous solid solution [39]. Based on the pseudo-binary CeCo$_5$-CeCu$_5$ phase diagram by Girodin *et al.* [22], the segregation in the high $H_c$ sample is compatible with a classical precipitation hardening procedure [39]: after homogenization and quenching, the Ce(Co$_{0.8}$Cu$_{0.2}$)$_{5.4}$ solution decomposes during ageing within the miscibility gap into Co-rich and Cu-rich CaCu$_5$-type phases, forming the cellular nanostructure. The decomposition within the miscibility gap occurs via spinodal decomposition rather than through nucleation and growth, as the observed cellular structure is regular and nearly periodic [39]. In the low $H_c$ "homogenized" sample, a residual segregation likely results from sluggish Co/Cu diffusion, as Girodin *et al.* [22] note that full homogenization requires ~15 days, much longer than the 2h homogenization treatment in our study. This indicates that Cu segregation is thermally stable, likely contributing to the thermal robustness of the magnetic hardness of the material.

*(3) Coercivity mechanism*: Magnetometry reveals a pinning-type coercivity mechanism in Ce(Co$_{0.8}$Cu$_{0.2}$)$_{5.4}$. TEM magnetic imaging and micromagnetic simulations show that its nanostructure governs DW pinning, with the zig-zag domain pattern following the underlying nanoscale features (see point 4 below). Cu concentration variation $x$ within the Ce(Co$_{1-x}$Cu$_x$)$_{5.4}$ matrix microscopically affects local intrinsic properties, namely the magnetocrystalline anisotropy constant $K_u$ and the exchange constant $A$, reducing Bloch DW energy $\sigma_{dw}^B \sim \sqrt{AK_u}$ at higher Cu. This creates a spatially varying DW energy landscape $\sigma_{dw}^B(r)$, with lower energy in Cu-rich cell boundaries and higher energy in Cu-poor cells, causing attractive pinning at the cell boundaries. While both $K_u$ and $A$ decrease with increasing Cu, $K_u$ is roughly ~10 times more sensitive to Cu changes [22,25], justifying the Cu-independent $A$ in the simulations.

Generally, effective pinning occurs when the dimensions of the pinning cites match the domain wall width [38]. This criterion is fulfilled in Ce(Co$_{0.8}$Cu$_{0.2}$)$_{5.4}$, as the domain wall width of 4.3 nm, determined by off-axis EH, closely matches the ~5 nm width of the Cu-rich cell boundaries measured by APT, explaining the significant coercivity of 1 T. In addition to the "*chemical* contribution" to pinning, the proposed structural ordering of Cu in the Ce(Co,Cu)$_5$ phase of the Cu-rich cell boundaries may further add to the pinning strength as a "*structural* contribution", but its role remains to be quantified through *ab initio* calculations and structural characterization.

The higher Cu-gradient $\frac{dx}{dr}$ in the high $H_c$ samples compared to the low $H_c$ sample amplifies the spatial gradient of domain wall energy $\frac{d\sigma_{dw}^B}{dr}$, thus enhancing pinning strength and coercivity, supported by micromagnetic simulations. The higher pinning strength is further evidenced by the initial curves in **Figure 1e**, where the high $H_c$ sample exhibits a higher depinning field.

Micromagnetic models generally tend to overestimate coercivity [40,41], though underestimation occurs in some cases such as FePt/FeRh bilayers [42]. Underestimation here may stem from approximations such as pre-nucleated reversal domains, neglected *structural* contributions to the pinning affecting $K_u$ and $A$ in the Cu – ordered Ce(Co,Cu)$_5$ cell boundary phase, and generally underestimated $K_u$ and $A$ values of Ce(Co$_{1-x}$Cu$_x$)$_{5.4}$. For example, the value A = 2.74 pJm$^{-1}$ obtained from the experimental domain wall width of 4.3 nm slightly deviates from $A \approx 5.4$ pJm$^{-1}$, calculated from the formula from in [25] and $T_c$ values of Ce(Co$_{0.8}$Cu$_{0.2}$)$_5$ from [22], assuming a distance between two Co-atoms of 0.3 nm. Similarly, $K_u$ values used in this study represent a rough estimate, as they are volume-averaged across locally inhomogeneous samples, as demonstrated by the "homogenized" low $H_c$ sample. Independent determination of $K_u$ and $A$ are hence suggested. Furthermore, whether the Cu-rich boundaries are ferromagnetic or paramagnetic remains unclear, possibly affecting pinning.

*(4) Comparison to related systems*: The nanostructure in Ce(Co$_{0.8}$Cu$_{0.2}$)$_{5.4}$ bears some resemblance to that in pinning-type Sm$_2$Co$_{17}$-type magnets: (a) Cu-poor cells are surrounded by Cu-rich cell boundaries, consisting of regular and Cu-ordered Ce(Co,Cu)$_5$ phases for the former and Sm$_2$Co$_{17}$ and SmCo$_5$ phases for the latter [43–45]; (b) Both structures are elongated and tilted: in Ce(Co$_{0.8}$Cu$_{0.2}$)$_{5.4}$, the cellular structures are inclined up to 20° to the c-axis, while in Sm$_2$Co$_{17}$-type magnets, the Cu-rich cell boundaries form along $(01\bar{1}1)$ planes of the matrix phase, corresponding to an inclination of ~30° [46]; (c) High coercivity is similarly attributed to the high-Cu-gradient induced pinning [5,47,48]; (d) The "structural contrast" contributes to pinning in addition to the chemical Cu-gradient: in Ce(Co$_{0.8}$Cu$_{0.2}$)$_{5.4}$, the structure likely transitions from a regular 1:5 phase within the cells to a to Cu-ordered 1:5 phase at the cell boundaries, whereas in Sm$_2$Co$_{17}$-type magnets, it transitions from the 2:17 phase in the cells to the 1:5 phase at the cell boundaries [43–45]; (e) Domain structures are slightly similar: mesoscale MFM shows a fine out-of-plane domain structure for both systems, characteristic of pinning-type magnets containing a cellular structure [36]; nanoscale TEM Lorentz imaging shows a zig-zag-like in-plane domain structure, following Cu-rich cell boundaries for both systems, though less regular and interrupted in Ce(Co$_{0.8}$Cu$_{0.2}$)$_{5.4}$ [49,50].

At the same time, differences emerge when comparing the two systems: (a) Z-phase platelets as a third phase, found in 2:17 magnets perpendicular to the c-axis [43–45], are absent in Ce(Co$_{0.8}$Cu$_{0.2}$)$_{5.4}$; (b) Ce(Co$_{0.8}$Cu$_{0.2}$)$_{5.4}$ has considerably finer structure with ~5-10 nm large cells and 5 nm thin boundaries, compared to ~100 nm large cells and ~10 nm thin cell boundaries and Z-platelets in the 2:17 system [43–45]; (c) In general, Sm$_2$Co$_{17}$-type magnets feature three structurally and chemically distinct ferromagnetic phases [43–45], whereas Ce(Co$_{0.8}$Cu$_{0.2}$)$_{5.4}$ shows only two chemically different, but structurally closely related phases: a Cu-poor ferromagnetic phase with Co-content >70 at. % and a Cu-rich phase that is likely ferromagnetic when Cu < 40 at.% and paramagnetic when Cu >40 at.% [22].

Thus, pinning in Ce(Co$_{0.8}$Cu$_{0.2}$)$_{5.4}$ is a variant of conventional pinning, yet slightly different due a finer scale and lower structural and chemical contrast of the underlying nanostructure.

Critically, these results provide a plausible structural explanation for the long-standing concept of "giant intrinsic magnetic hardness" in Ce(Co,Cu)$_5$ and related systems such as SmCo$_{5-x}$M$_x$ (M= Cu, Al, Ni etc.) [17–20], a mechanism that has remained unclear since the 1970s. Once thought to be homogenous and single phase, the magnetic hardness in these systems was considered fundamentally different from precipitation-hardened magnets, possibly stemming from the periodicity of the crystal lattice. However, our findings suggest otherwise: the *chemical* contrast, possibly combined with subtle *structural* modifications such as ordering of Cu in the 1:5 phase, enables effective DW pinning, even in nearly single-phase Ce(Co$_{0.8}$Cu$_{0.2}$)$_{5.4}$ hard magnets. The nanoscale segregation and structural ordering revealed here by a combination of APT and TEM likely went undetected in earlier studies due to the small feature size of the nanostructure and the low Co/Cu contrast in conventional techniques such as TEM-EDX and XRD, where the signal scales with the atomic number. In contrast, the mass-spectrum-based APT clearly distinguishes between Co and Cu [34], uncovering the segregation. These findings likely extend to other systems exhibiting giant intrinsic magnetic hardness such as YCo$_{5-x}$Ni$_x$, ThCo$_{5-x}$Ni, SmCo$_{3-x}$Ni, SmCo$_{2-x}$Ni$_x$, and Sm$_2$Co$_{17-x}$Al$_x$ [17–20].

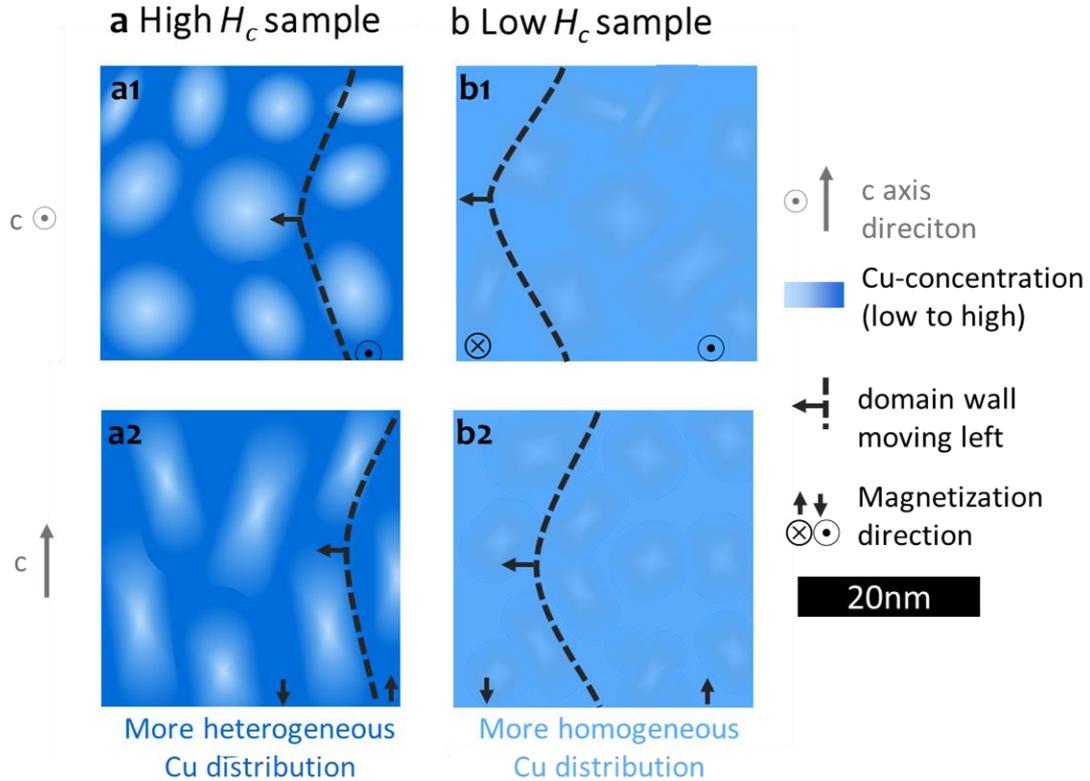

**Figure 8 Schematic representations of the coercivity mechanism in Ce(Co$_{0.8}$Cu$_{0.2}$)$_{5.4}$. (a)** High $H_c$ sample: pronounced chemical contrast and preferred orientation of Cu-poor cylindrical cells leads to strong pinning at Cu-rich cell boundaries. **(b)** Low $H_c$ sample: weaker compositional contrast and random cell orientation reduce pinning effectiveness. The DW (dashed line) propagates more easily through the less segregated structure.

# 5 Conclusion

Our study demonstrates that the coercivity in Ce(Co$_{0.8}$Cu$_{0.2}$)$_{5.4}$ originates from chemical segregation and structural ordering rather than a homogeneous single-phase state. Despite similar mesoscale microstructures, APT and TEM revealed clear nanoscale differences between heat-treated samples: the aged high $H_c$ sample shows Cu-poor ferromagnetic cylindrical cells, separated by Cu-rich cell boundaries with sharp compositional gradients, whereas the homogenized low $H_c$ sample exhibits a more random structure with weaker segregation. The cells exhibit a regular Ce(Co,Cu)$_5$ phase with random Cu substitution on Co-sites, whereas the cell boundaries likely host a structurally modified Ce(Co,Cu)$_5$ phase with Cu-ordering, forming coherent interfaces between the cells and the boundaries. The presence of chemical segregation is due to spinodal decomposition.

Magnetization reversal is of the pinning type, and micromagnetic simulations and TEM-based magnetic imaging confirm that the cellular nanostructure provides effective pinning centers for domain walls. The chemical modulation of Cu content creates an energy landscape for domain walls, leading to effective pinning at Cu-rich boundaries. The gradient in magnetocrystalline anisotropy across these interfaces, likely enhanced by structural ordering, plays a dominant role in determining coercivity. Comparison to $Sm_2Co_{17}$-type magnets highlights common features such as Cu-gradient-driven pinning and zigzag-like domain structures, and shows that $Ce(Co_{0.8}Cu_{0.2})_{5.4}$ exhibits a finer-scale variation of conventional pinning with lower structural and chemical contrast in its underlying nanostructure.

Importantly, the discovery of nanoscale chemical segregation in nearly single-phase $Ce(Co_{0.8}Cu_{0.2})_{5.4}$ magnets offers a microstructural explanation for the long-standing phenomenon of "giant intrinsic magnetic hardness" in systems such as $SmCo_{5-x}M_x$, $YCo_{5-x}Ni_x$, $ThCo_{5-x}Ni$, $SmCo_{3-x}Ni$, $SmCo_{2-x}Ni_x$, and $Sm_2Co_{17-x}Al_x$. These findings open up new possibilities for designing next-generation rare-earth-lean permanent magnets by leveraging controlled nanoscale segregation.

# 6 Acknowledgements

We acknowledge funding by the Deutsche Forschungsgemeinschaft (DFG, German Research Foundation), Project ID No. 405553726-CRC/TRR 270 DFG and by the German BMBF under the grant number 03XP0166A. NP is grateful for the funding for his scholarship by the IMPRS SURMAT. NP and BG are grateful for the funding of the Leibniz Prize 2020 by the DFG. NP and BG thank Uwe Tezins, Christian Broß, and Andreas Sturm for their support to the FIB & APT facilities at MPI-SusMat. L. M.-L acknowledges the European Research Council (ERC) "Horizon 2020" Program under Grant No. 805359-FOXON and Grant No. 957521-STARE. Authors Y.Y., and B.-X.X. appreciate their access to the Lichtenberg High-Performance Computer and the technique supports from the HHLR, Technical University of Darmstadt. Y.Y. also highly thanks Dr. Jiajun Sun and the research assistant Guanyu Chen for performing intensive micromagnetic simulations.

# Literature

[1] O. Gutfleisch, M.A. Willard, E. Brück, C.H. Chen, S.G. Sankar, J.P. Liu, Magnetic materials and devices for the 21st century: Stronger, lighter, and more energy efficient, Adv. Mater. 23 (2011) 821–842. https://doi.org/10.1002/adma.201002180.

[2] S. Bobba, S. Carrara, J. Huisman, F. Mathieux, C. Pavel, Critical raw materials for strategic technologies and sectors in the EU – A foresight study, European Commission: Directorate-General for Internal Market, Industry, Entrepreneurship and SMEs, 2020. https://doi.org/doi/10.2873/58081.

[3] E.A. Nesbitt, New Permanent Magnet Materials Containing Rare-Earth Metals, J. Appl. Phys. 40 (1969) 1259–1265. https://doi.org/10.1063/1.1657619.

[4] H. Wang, D. Zhang, Y. Li, J. Yang, W. Song, W. Liu, M. Yue, Deformation and texture formation mechanism of hot-deformed SmCo5 nanocrystalline magnets, J. Alloys Compd. 934 (2023) 168071. https://doi.org/10.1016/j.jallcom.2022.168071.

[5] F. Staab, E. Bruder, L. Schäfer, K. Skokov, D. Koch, B. Zingsem, E. Adabifiroozjaei, L. Molina-Luna, O. Gutfleisch, K. Durst, Hard magnetic SmCo5-Cu nanocomposites produced by severe plastic deformation, Acta Mater. 246 (2023) 118709. https://doi.org/10.1016/j.actamat.2023.118709.

[6] Z. Yang, Y. Wang, H. Xu, Q. Wu, H. Zhang, W. Liu, M. Yue, Strategies for the Synthesis of Nanostructured SmCo 5 Magnetic Particles for Permanent Magnetic Application, ACS Appl. Nano Mater. 7 (2024) 4252–4263. https://doi.org/10.1021/acsanm.3c05776.

[7] X. Ye, H.K. Singh, H. Zhang, H. Geßwein, M.R. Chellali, R. Witte, A. Molinari, K. Skokov, O. Gutfleisch, H. Hahn, R. Kruk, Giant voltage-induced modification of magnetism in micron-scale ferromagnetic metals by hydrogen charging, Nat. Commun. 11 (2020) 4849. https://doi.org/10.1038/s41467-020-18552-z.

[8] H. Zhang, A. Aubert, F. Maccari, C. Dietz, M. Yue, O. Gutfleisch, K. Skokov, Study of magnetization reversal and magnetic hardening in SmCo5 single crystal magnets, J. Alloys Compd. 993 (2024) 174570. https://doi.org/10.1016/j.jallcom.2024.174570.

[9] W. Tang, Y. Zhang, G.C. Hadjipanayis, Effect of Zr on the microstructure and magnetic properties of Sm(CobalFe0.1Cu0.088Zrx)8.5 magnets, J. Appl. Phys. 87 (2000) 399–403. https://doi.org/10.1063/1.371874.

[10] W. Tang, Y. Zhang, G.C. Hadjipanayis, Microstructure, magnetic properties and magnetic hardening in 2:17 Sm-Co magnets, Zeitschrift Fuer Met. Res. Adv. Tech. 93 (2002) 1002–1008. https://doi.org/10.3139/146.021002.

[11] Y. Zhang, W. Tang, G.C. Hadjipanayis, C. Chen, C. Nelson, K. Krishnan, Evolution of microstructure, microchemistry and coercivity in 2 : 17 type Sm-Co magnets with heat treatment, IEEE Trans. Magn. 37 (2001) 2525–2527. https://doi.org/10.1109/20.951223.

[12] W. Tang, Y. Zhang, G.C. Hadjipanayis, H. Kronmüller, Influence of Zr and Cu content on the microstructure and coercivity in Sm(CobalFe0.1CuyZrx)8.5 magnets, J. Appl. Phys. 87 (2000) 5308–5310. https://doi.org/10.1063/1.373330.

[13] N. Polin, K.P. Skokov, A. Aubert, H. Zhang, B. Ekitli, E. Adabifiroozjaei, L. Molina-Luna, O. Gutfleisch, B. Gault, Formation of cellular/lamellar nanostructure in Sm2Co17-type binary and ternary Sm-Co-Zr magnets, Scr. Mater. 258 (2025) 116530. https://doi.org/10.1016/j.scriptamat.2024.116530.

[14] N.T. Nassar, X. Du, T.E. Graedel, Criticality of the Rare Earth Elements, J. Ind. Ecol. 19 (2015) 1044–1054. https://doi.org/10.1111/jiec.12237.

[15] W.A.J.J. Velge, K.H.J. Buschow, Magnetic and crystallographic properties of some rare earth cobalt compounds with CaZn5 structure, J. Appl. Phys. 39 (1968) 1717–1720. https://doi.org/10.1063/1.1656420.

[16] S.M. Kim, W.J.L. Buyers, H. Lin, E. Bauer, Structure of the heavy electron compounds Ce(Cu x Al1-x )5 and Ce(Cu x Ga1-x )5, [0.6?x?0.8], Zeitschrift F�r Phys. B Condens. Matter 84 (1991) 201–203. https://doi.org/10.1007/BF01313537.

[17] J. van den Broek, H. Zijlstra, Calculation of intrinsic coercivity of magnetic domain walls in perfect crystals, IEEE Trans. Magn. 7 (1971) 226–230. https://doi.org/10.1109/TMAG.1971.1067036.

[18] F.F. Westendorp, Domains in SmCo5 at Low Temperatures, Appl. Phys. Lett. 20 (1972) 441–443. https://doi.org/10.1063/1.1654008.

[19] H.R. Hilzinger, H. Kronmüller, Spin Configuration and Intrinsic Coercive Field of Narrow Domain Walls in Co 5 R-Compounds, Phys. Status Solidi 54 (1972) 593–604. https://doi.org/10.1002/pssb.2220540223.

[20] H. Oesterreicher, Giant intrinsic magnetic hardness, Appl. Phys. 15 (1978) 341–354. https://doi.org/10.1007/BF00886151.

[21] H.J. Leamy, M.L. Green, The Structure of Co-Cu-Fe-Ce Permanent Magnets, IEEE Trans. Magn. 9 (1973) 205–209. https://doi.org/10.1109/TMAG.1973.1067642.

[22] D. Girodin, C.H. Allibert, F. Givord, R. Lemaire, PHASE EQUILIBRIA IN THE CeCo5-CeCu5 SYSTEM AND Structural Chracterization of the Ce(Co_1-x-Cu_x) phases, J. Less-Common Met. 110 (1985) 149–158. https://doi.org/10.1016/0022-5088(85)90316-9.

[23] O. Palasyuk, M. Onyszczak, T.H. Kim, L. Zhou, M.J. Kramer, S.L. Bud'ko, P.C. Canfield, A. Palasyuk, Structural and magnetic properties of hard magnetic system Ce(Co1-xFex)4.4Cu0.6 (0 ≤ x ≤ 0.19), J. Alloys Compd. 883 (2021). https://doi.org/10.1016/j.jallcom.2021.160866.

[24] W. Sun, K. Song, M. Zhu, Y. Fang, N. Yu, S. Wang, W. Li, Effect of the Ce Content

on the Magnetic Properties and Microstructure of CeCo5-based Sintered Bulk Magnets, J. Supercond. Nov. Magn. 31 (2018) 1761–1765. https://doi.org/10.1007/s10948-017-4387-8.

[25] E. Lectard, C.H. Allibert, R. Ballou, Saturation magnetization and anisotropy fields in the Sm(Co 1– x Cu x ) 5 phases, J. Appl. Phys. 75 (1994) 6277–6279. https://doi.org/10.1063/1.355423.

[26] K. Thompson, D. Lawrence, D.J. Larson, J.D. Olson, T.F. Kelly, B. Gorman, In situ site-specific specimen preparation for atom probe tomography, Ultramicroscopy 107 (2007) 131–139. https://doi.org/10.1016/J.ULTRAMIC.2006.06.008.

[27] A. Vansteenkiste, J. Leliaert, M. Dvornik, M. Helsen, F. Garcia-Sanchez, B. Van Waeyenberge, The design and verification of MuMax3, AIP Adv. 4 (2014). https://doi.org/10.1063/1.4899186.

[28] L. Exl, S. Bance, F. Reichel, T. Schrefl, H. Peter Stimming, N.J. Mauser, LaBonte's method revisited: An effective steepest descent method for micromagnetic energy minimization, J. Appl. Phys. 115 (2014). https://doi.org/10.1063/1.4862839.

[29] A. Furuya, J. Fujisaki, K. Shimizu, Y. Uehara, T. Ataka, T. Tanaka, H. Oshima, Semi-Implicit Steepest Descent Method for Energy Minimization and Its Application to Micromagnetic Simulation of Permanent Magnets, IEEE Trans. Magn. 51 (2015) 1–4. https://doi.org/10.1109/TMAG.2015.2439290.

[30] R.D. McMichael, M.J. Donahue, D.G. Porter, J. Eicke, Comparison of magnetostatic field calculation methods on two-dimensional square grids as applied to a micromagnetic standard problem, J. Appl. Phys. 85 (1999) 5816–5818. https://doi.org/10.1063/1.369929.

[31] H. Fangohr, G. Bordignon, M. Franchin, A. Knittel, P.A.J. de Groot, T. Fischbacher, A new approach to (quasi) periodic boundary conditions in micromagnetics: The macrogeometry, J. Appl. Phys. 105 (2009). https://doi.org/10.1063/1.3068637.

[32] M.P. Moody, L.T. Stephenson, A. V. Ceguerra, S.P. Ringer, Quantitative binomial distribution analyses of nanoscale like-solute atom clustering and segregation in atom probe tomography data, Microsc. Res. Tech. 71 (2008) 542–550. https://doi.org/10.1002/jemt.20582.

[33] V.J. Araullo-Peters, B. Gault, S.L. Shrestha, L. Yao, M.P. Moody, S.P. Ringer, J.M. Cairney, Atom probe crystallography: Atomic-scale 3-D orientation mapping, Scr. Mater. 66 (2012) 907–910. https://doi.org/10.1016/j.scriptamat.2012.02.022.

[34] B. Gault, M.P. Moody, J.M. Cairney, S.P. Ringer, Atom Probe Microscopy, Springer New York, New York, NY, 2012. https://doi.org/10.1007/978-1-4614-3436-8.

[35] B.P. Geiser, T.F. Kelly, D.J. Larson, J. Schneir, J.P. Roberts, Spatial Distribution Maps for Atom Probe Tomography, Microsc. Microanal. 13 (2007) 437–447. https://doi.org/10.1017/S1431927607070948.

[36] O. Gutfleisch, K.-H.H. Müller, K. Khlopkov, M. Wolf, A. Yan, R. Schäfer, T. Gemming, L. Schultz, Evolution of magnetic domain structures and coercivity in high-performance SmCo 2:17-type permanent magnets, Acta Mater. 54 (2006) 997–1008. https://doi.org/10.1016/j.actamat.2005.10.026.

[37] A. Kovács, R.E. Dunin-Borkowski, Magnetic Imaging of Nanostructures Using Off-Axis Electron Holography, in: 2018: pp. 59–153. https://doi.org/10.1016/bs.hmm.2018.09.001.

[38] J.M.D. Coey, Magnetism and Magnetic Materials, Cambridge University Press, 2001. https://doi.org/10.1017/CBO9780511845000.

[39] D.A. Porter, K.E. Easterling, M.Y. Sherif, Phase Transformations in Metals and Alloys, CRC Press, Boca Raton, 2021. https://doi.org/10.1201/9781003011804.

[40] C. Bhandari, G. Nop, J. Smith, D. Paudyal, Accurate machine-learning predictions of coercivity in high-performance permanent magnets, Phys. Rev. Appl. (2023). https://doi.org/10.1103/physrevapplied.22.024046.

[41] G. Zhao, X. Wang, Nucleation, pinning, and coercivity in magnetic nanosystems : An analytical micromagnetic approach, Phys. Rev. B 74 (2006) 12409. https://doi.org/10.1103/PHYSREVB.74.012409.

[42] F. García-Sánchez, O. Chubykalo-Fesenko, O. Mryasov, R. Chantrell, K. Guslienko, Exchange spring structures and coercivity reduction in FePt∕FeRh bilayers: A comparison of multiscale and micromagnetic calculations, Appl. Phys. Lett. 87 (2005) 122501. https://doi.org/10.1063/1.2051789.

[43] O. Gutfleisch, High-Temperature Samarium Cobalt Permanent Magnets, in: Nanoscale Magn. Mater. Appl., Springer US, Boston, MA, 2009: pp. 337–372. https://doi.org/10.1007/978-0-387-85600-1_12.

[44] F. Okabe, H.S. Park, D. Shindo, Y.-G. Park, K. Ohashi, Y. Tawara, Microstructures and Magnetic Domain Structures of Sintered Sm(Co$_{0.720}$Fe$_{0.200}$Cu$_{0.055}$Zr$_{0.025}$)$_{7.5}$ Permanent Magnet Studied by Transmission Electron Microscopy, Mater. Trans. 47 (2006) 218–223. https://doi.org/10.2320/matertrans.47.218.

[45] S. Giron, N. Polin, E. Adabifiroozjaei, Y. Yang, A. Kovács, T.P. Almeida, D. Ohmer, K. Uestuener, A. Saxena, M. Katter, others, Towards engineering the perfect defect in high-performing permanent magnets, ArXiv Prepr. ArXiv2304.14958 (2023). https://doi.org//10.48550/arXiv.2304.14958.

[46] H. Wu, C. Zhang, Z. Liu, G. Wang, H. Lu, G. Chen, Y. Li, R. Chen, A. Yan, Nanoscale short-range ordering induced cellular structure and microchemistry evolution in Sm2Co17-type magnets, Acta Mater. 200 (2020) 883–892. https://doi.org/10.1016/j.actamat.2020.09.057.

[47] S. Sharma, A. Zintler, D. Günzing, J. Lill, D.M. Meira, R. Eilhardt, H.K. Singh, R. Xie, G. Gkouzia, M. Major, I. Radulov, P. Komissinskiy, H. Zhang, K. Skokov, H. Wende, Y.K. Takahashi, K. Ollefs, L. Molina-Luna, L. Alff, Epitaxy Induced Highly Ordered Sm 2 Co 17 –SmCo 5 Nanoscale Thin-Film Magnets, ACS Appl. Mater. Interfaces 13 (2021) 32415–32423. https://doi.org/10.1021/acsami.1c04780.

[48] G. Gkouzia, D. Günzing, R. Xie, T. Weßels, A. Kovács, A.T. N'Diaye, M. Major, J.P. Palakkal, R.E. Dunin-Borkowski, H. Wende, H. Zhang, K. Ollefs, L. Alff, Element-Specific Study of Magnetic Anisotropy and Hardening in SmCo5-xCux Thin Films, Inorg. Chem. 62 (2023) 16354–16361. https://doi.org/10.1021/acs.inorgchem.3c01768.

[49] J. Fidler, J. Bernardi, P. Skalicky, Analytical Electron Microscope Study of High- and Low-Coercivity SmCo 2:17 Magnets, MRS Online Proc. Libr. 96 (2011) 181. https://doi.org/10.1557/PROC-96-181.

[50] Y. Zhang, W. Tang, G.C. Hadjipanayis, C.H. Chen, D. Goll, H. Kronmuller, Magnetic domain structure in SmCo 2:17 permanent magnets, IEEE Trans. Magn. 39 (2003) 2905–2907. https://doi.org/10.1109/TMAG.2003.815743.

[51] A. Saxena, N. Polin, N. Kusampudi, S. Katnagallu, L. Molina-Luna, O. Gutfleisch,

B. Berkels, B. Gault, J. Neugebauer, C. Freysoldt, A Machine Learning Framework for Quantifying Chemical Segregation and Microstructural Features in Atom Probe Tomography Data, Microsc. Microanal. 29 (2023) 1658–1670. https://doi.org/10.1093/micmic/ozad086.

[52] T.N. Lamichhane, M.T. Onyszczak, O. Palasyuk, S. Sharikadze, T.H. Kim, Q. Lin, M.J. Kramer, R.W. McCallum, A.L. Wysocki, M.C. Nguyen, V.P. Antropov, T. Pandey, D. Parker, S.L. Bud'ko, P.C. Canfield, A. Palasyuk, Single-Crystal Permanent Magnets: Extraordinary Magnetic Behavior in the Ta -, Cu -, and Fe -Substituted CeCo5 Systems, Phys. Rev. Appl. 11 (2019) 1. https://doi.org/10.1103/PhysRevApplied.11.014052.

# Supplementary information:
# Direct observation of nanoscale pinning centers in quasi single-phase Ce(Co$_{0.8}$Cu$_{0.2}$)$_{5.4}$ magnets

Nikita Polin[1], Shangbin Shen[2], Fernando Maccari[2], Alex Aubert[2], Esmaeil Adabifiroozjaei[3], Tatiana Smoliarova[5], Yangyiwei Yang[2], Xinren Chen[1], Yurii Skourski[6], Alaukik Saxena[1], András Kovács[7], Rafal E. Dunin-Borkowski[7], Michael Farle[5], Bai-Xiang Xu[2], Leopoldo Molina-Luna[3], Oliver Gutfleisch[2], Baptiste Gault[1,4], Konstantin Skokov[2]

[1] Max-Planck-Institut für Nachhaltige Materialien, 40237 Düsseldorf, Germany
[2] Institute of Materials Science, Technische Universität Darmstadt, 64287 Darmstadt, Germany
[3] Advanced Electron Microscopy Division, Institute of Materials Science, Department of Materials and Geosciences, Technische Universität Darmstadt, Peter-Grünberg-Str. 2, Darmstadt 64287, Germany
[4] Department of Materials, Royal School of Mines, Imperial College, Prince Consort Road, London SW7 2BP, United Kingdom
[5] Faculty of Physics and Center for Nanointegration (CENIDE), University Duisburg Essen, Duisburg 47057, Germany
[6] Dresden High Magnetic Field Laboratory (HLD-EMFL), Helmholtz-Zentrum Dresden-Rossendorf, Dresden 01328, Germany
[7] Ernst Ruska-Centre for Microscopy and Spectroscopy with Electrons, Forschungszentrum Jülich, Jülich 52425, Germany

## A. Magnetic characterization

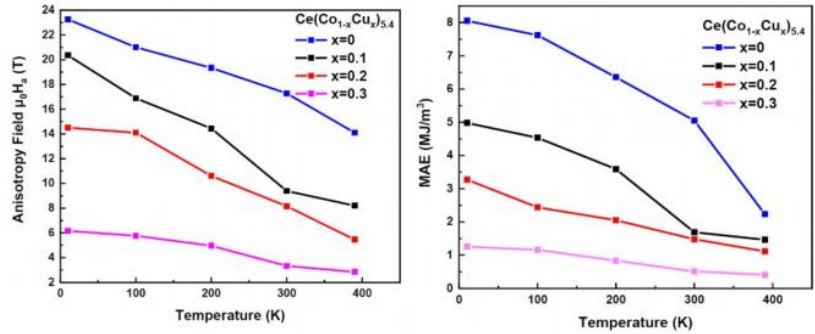

**Figure S1** Intrinsic properties of the series of Ce(Co$_{1-x}$Cu$_x$)$_{5.4}$. The anisotropy field (left) and anisotropy energy (right) experimentally determined from homogenized nearly single-phase grains at different temperatures.

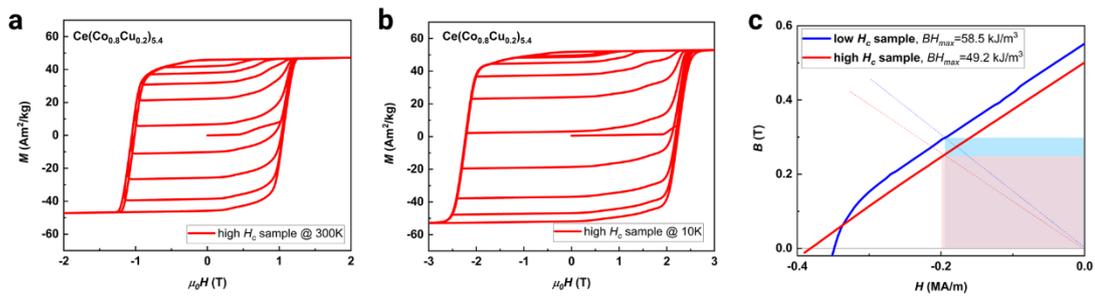

**Figure S2** Magnetometry analysis: (a-b) partial hysteresis loops of the high $H_c$ Ce(Co$_{0.8}$Cu$_{0.2}$)$_{5.4}$ sample at 300K (a) and 10K (b). (c) second quadrant of the *B-H* hysteresis loop with $BH_{max}$ indicated as rectangles. The maximum energy product $BH_{max}$ for the low and high $H_c$ samples are 58.5 kJ/m$^3$ and 49.2 kJ/m$^3$.

## B. Microstructure of polycrystalline samples

As Figure S3 shows, the microstructure of the high $H_c$ sample is characterized by large grains of ~100-500µm (BSE image Figure 5a) which have high degree of texture with misorientation angles between the grain are < 3° (EBSD, Figure 5b). The contrast in the BSE images stems from the slightly different orientation of the grains as it correlates with EBSD contrast. Note that this sample is randomly oriented as the EBSD color map shows.

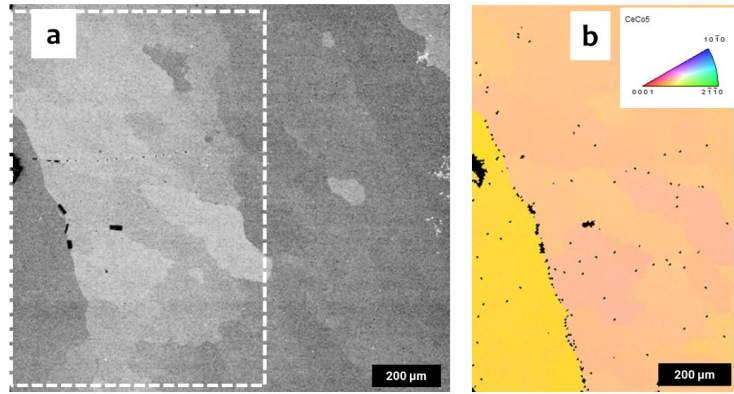

**Figure S3** Microstructure characterization of Ce(Co$_{0.8}$Cu$_{0.2}$)$_{5.4}$ high $H_c$ polycrystalline sample by electron microscopy (a) in backscatter electron mode (BSE) and by (b) electron backscatter diffraction (EBSD). The sample possesses large grains ~100-500 μm in size and a high degree of texture

## C. Atom probe tomography: specimen preparation and further analyses

**Table S1** Compositions and standard deviations derived from APT measurements shown in **Figure 3**. The calculations for Cu-rich and Cu-poor regions were executed as described in detail in Ref. [51]. The standard deviations for the total compositions are calculated from different APT measurements of the same sample.

| Sample | Region | Ce (at. %) | Co (at. %) | Cu (at. %) | Composition |
|---|---|---|---|---|---|
| High $H_c$ sample | Cu-rich | 15.77±0.18 | 63.62±0.37 | 20.61±0.32 | Ce(Co$_{0.76}$Cu$_{0.24}$)$_{5.34}$ |
| | Cu-poor | 16.80±0.22 | 74.98±0.46 | 8.22±0.26 | Ce(Co$_{0.90}$Cu$_{0.10}$)$_{4.95}$ |
| | Total | 16.19±0.17 | 69.47±0.69 | 14.35±0.15 | Ce(Co$_{0.83}$Cu$_{0.17}$)$_{5.18}$ |
| Low $H_c$ sample | Cu-rich | 15.57±0.35 | 63.31±0.40 | 21.13±0.67 | Ce(Co$_{0.75}$Cu$_{0.25}$)$_{5.42}$ |
| | Cu-poor | 16.39±0.31 | 72.19±0.44 | 11.42±0.22 | Ce(Co$_{0.86}$Cu$_{0.14}$)$_{5.10}$ |
| | Total | 15.96±0.16 | 68.93±0.69 | 15.11±0.15 | Ce(Co$_{0.82}$Cu$_{0.18}$)$_{5.27}$ |

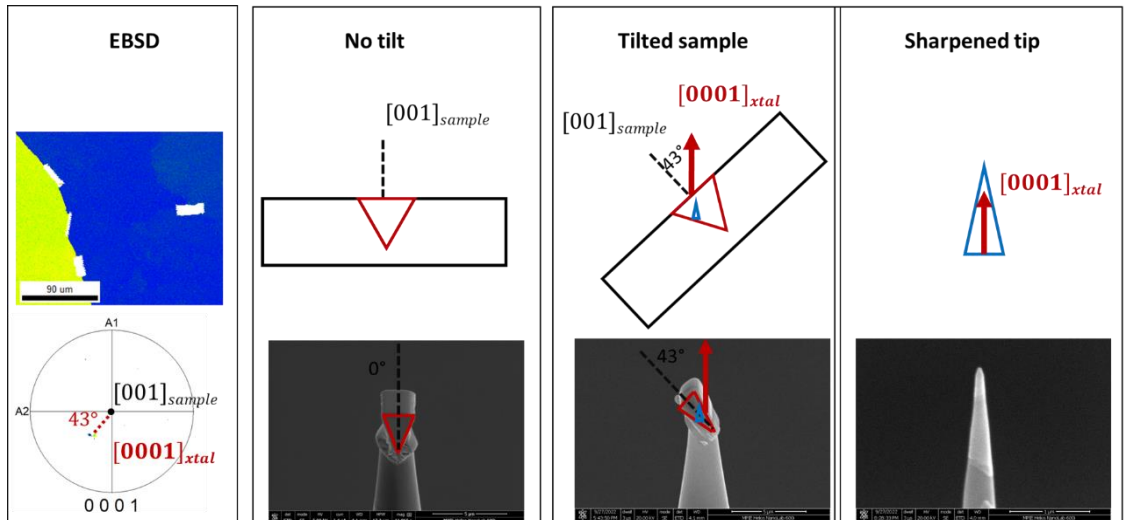

**Figure S4** APT needle preparation by FIB with a desired orientation. First image: EBSD map (top) and pole figure (bottom), determining the grain orientation to the sample's normal. Second image: the conventional lift-out schematics (top) and corresponding SEM image of the APT specimen before sharpening (bottom). Third image: schematics of the lift-out under an appropriate angle, prior determined by EBSD to be 43° (top), and corresponding SEM image of the APT specimen before sharpening (bottom). Forth image: sharpened APT tip with correct orientation schematics (top) and corresponding SEM image of the APT specimen after sharpening. Red and blue triangles indicate the APT samples before and after sharpening, the sample normal direction $[001]_{sample}$ and the *c*-axis of the crystal $[0001]_{xtal}$ enclose an angle of 43°.

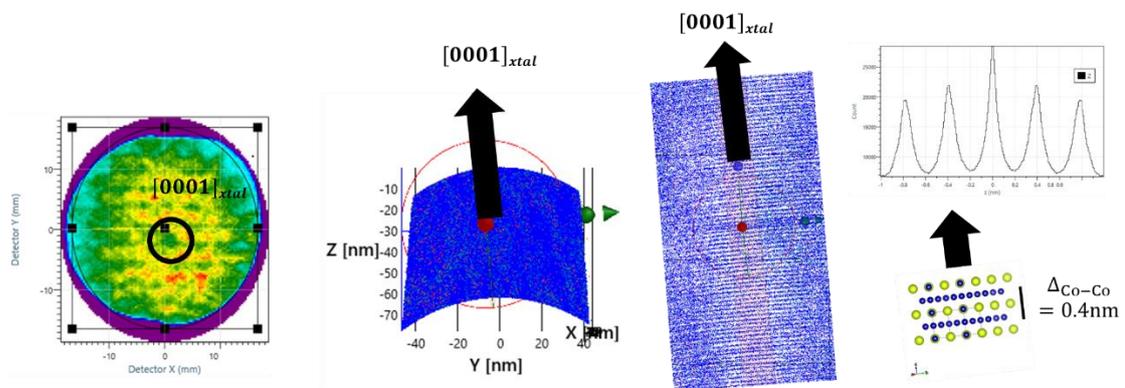

**Figure S5** APT crystallographic analysis. First: identification of the [0001] pole in the detector histogram. Second: low magnification APT reconstruction. Third: high magnification in a slice of an APT reconstruction with visible atomic planes. Fourth: spatial distribution map (SDM, top) and structure model (bottom).

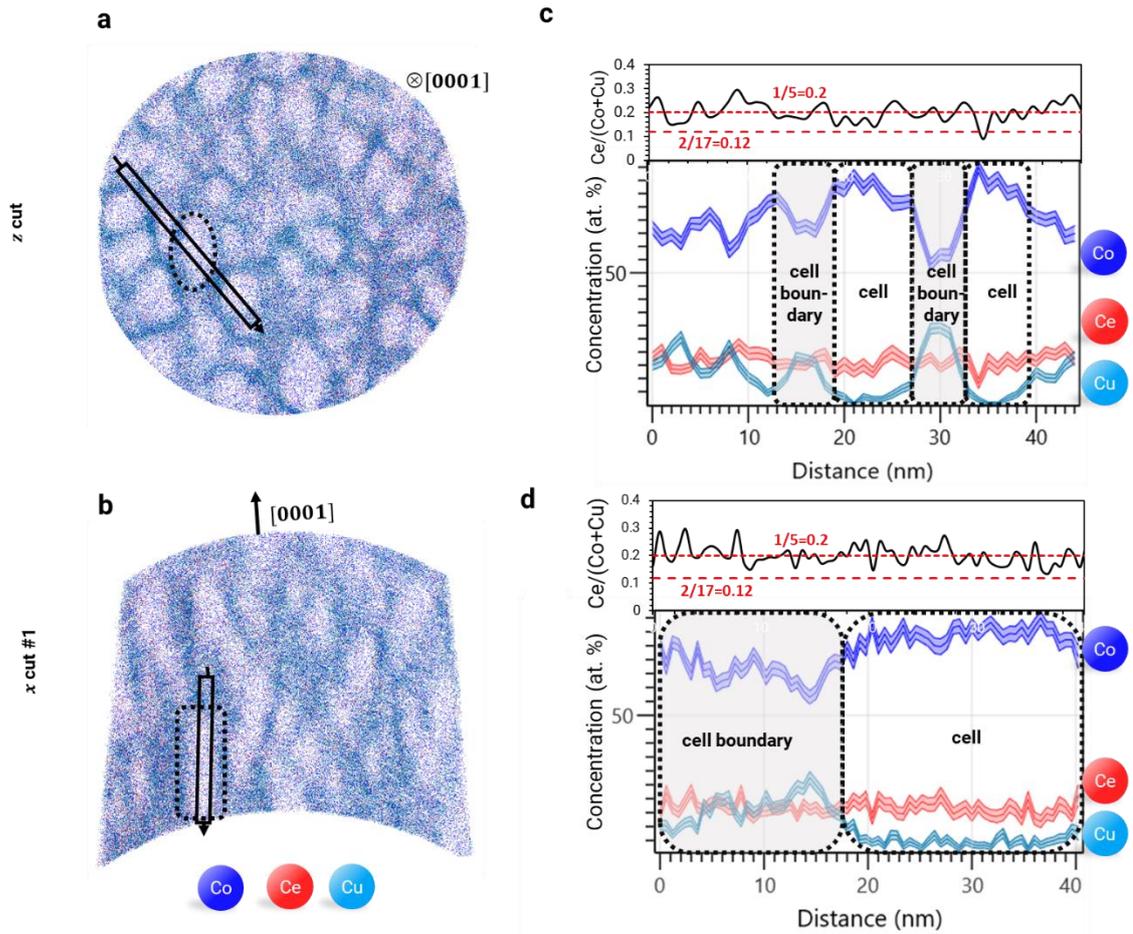

**Figure S6** APT analysis of the cellular structure in the high $H_c$ sample: (a,b) 3D reconstructions. (c,d) Additional 1D profiles and calculated rare earth/transition metals ratio Ce/(Co+Cu)..

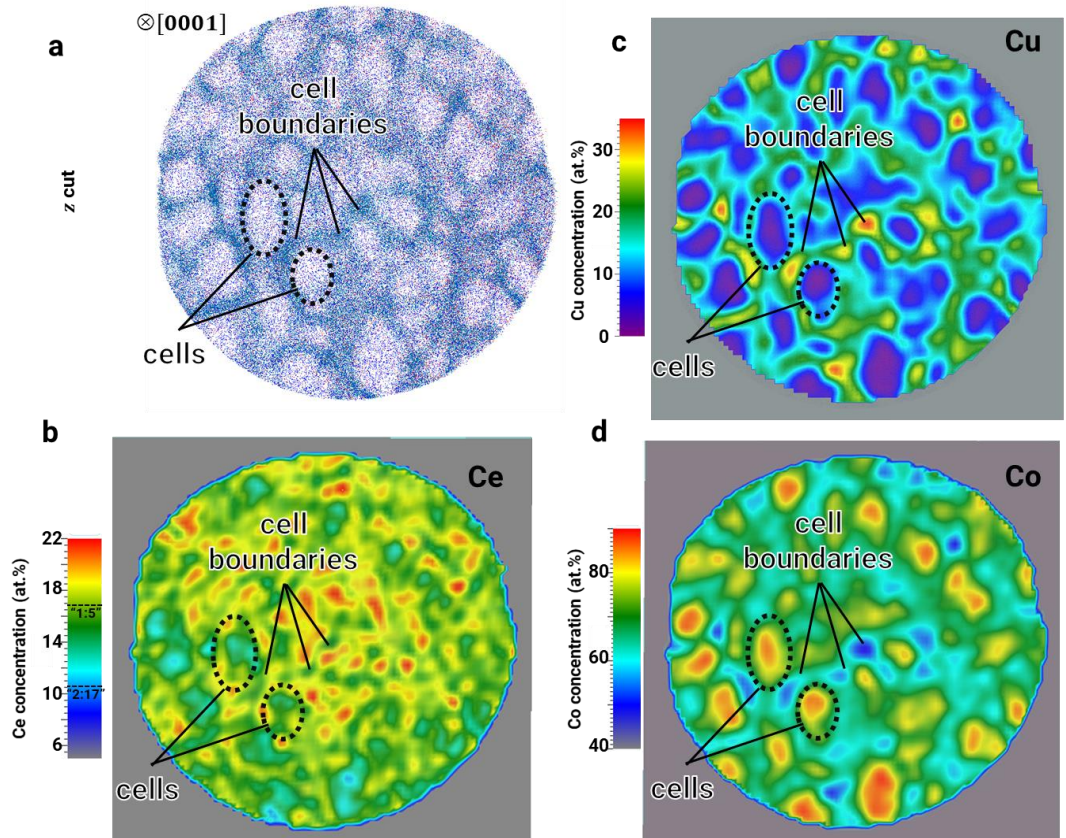

**Figure S7** APT analysis of the cellular structure in the high $H_c$ sample in cross section (z cut): (a) 10nm thin slice of APT data and corresponding 2D concentration maps of (b) Ce, (c) Cu and (d) Co. In the Ce colorbar, the Ce concentrations corresponding to the 1:5 (10.5 at.%) and 2:17 (16.7 at. %) phases are marked.

## D. Transmission electron microscopy: further analyses

Unlike the single-grain samples used in the remainder of this study, the TEM analysis in **Figure S9** was performed on a polycrystalline high $H_c$ sample to investigate grain boundary (GB) structure and chemistry.

In **Figure S9a-c**, bright-field STEM images with increasing magnification show a chemically homogeneous matrix, that is Cu-depleted relative to Cu-enriched grain boundaries (GBs), which contain ~100nm large Cu-rich precipitates, cf. EDX profile in **Figure S9d**. These GB features resemble those reported in Ta-containing Ce(Co,Cu)$_5$ single crystals with composition Ce$_{15.7}$Ta$_{0.6}$Co$_{67.8}$Cu$_{15.9}$, prepared by self-flux technique [52]. Based on low-magnification TEM image in **Figure S9**a, the precipitate volume fraction is roughly estimated be to <5%.

A detailed view of the GB precipitates in **Figure S9e** by STEM and corresponding energy-dispersive X-ray spectroscopy (EDX) chemical maps in **Figure S9f-h** reveal an increase in Cu, decrease of these GB precipitates. **Figure S9i** shows a high magnification BF TEM image of a representative precipitate with its associated fast Fourier transformed image (**Figure S9j**), displaying satellite peaks at $\frac{1}{2}\{110\}$. These features can be reproduced by simulations (**Figure S9l**), assuming an ordered arrangement of Cu atoms on Co sites along one of six symmetry-equivalent [100] directions, as depicted in **Figure S9**. Crystallographic data for the Ce(Co,Cu)$_5$ with Cu-ordering is available in the supplementary information (see CIF file).

Although the diffraction pattern also matches with Ce$_2$Co$_{17}$ phase as suggested by Leamy et al. [21], the expected compositional contrast between the CeCo$_5$ and Ce$_2$Co$_{17}$ phases is not observed in our data: EDX chemical map in **Figure S9h** and the EDX line profile in **Figure S9d** show no variation of Ce-content. This suggests, that no transition from 2:17 phase to 1:5 has occurred, in line with APT results from the matrix. Hence we conclude that in GB precipitates, higher Cu-concentration correlates to the presence of a Cu-ordered CeCo$_5$-phase.

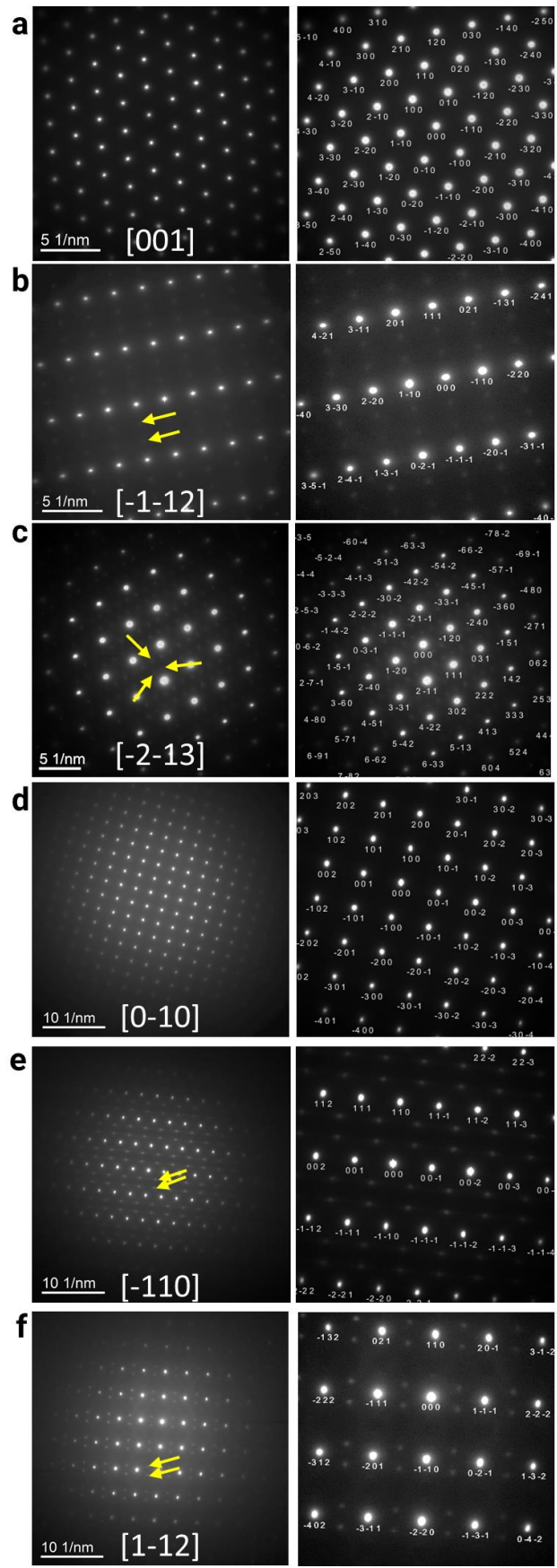

**Figure S8** Series of diffraction images from the matrix of the CeCoCu high $H_c$ sample with different zone axes indicated in the respective images, with unindexed pattern in the left row and indexed

main reflection in the right row. The extra reflections (satellite peaks) are indicated with yellow arrow and are only observed in higher index patters. Images (a-c) stem from a TEM specimen with c-axis out plane and (d-f) with c-axis in-plane. The satellite peaks are observed at orientations different from [001] (a) and [0-10] (d).

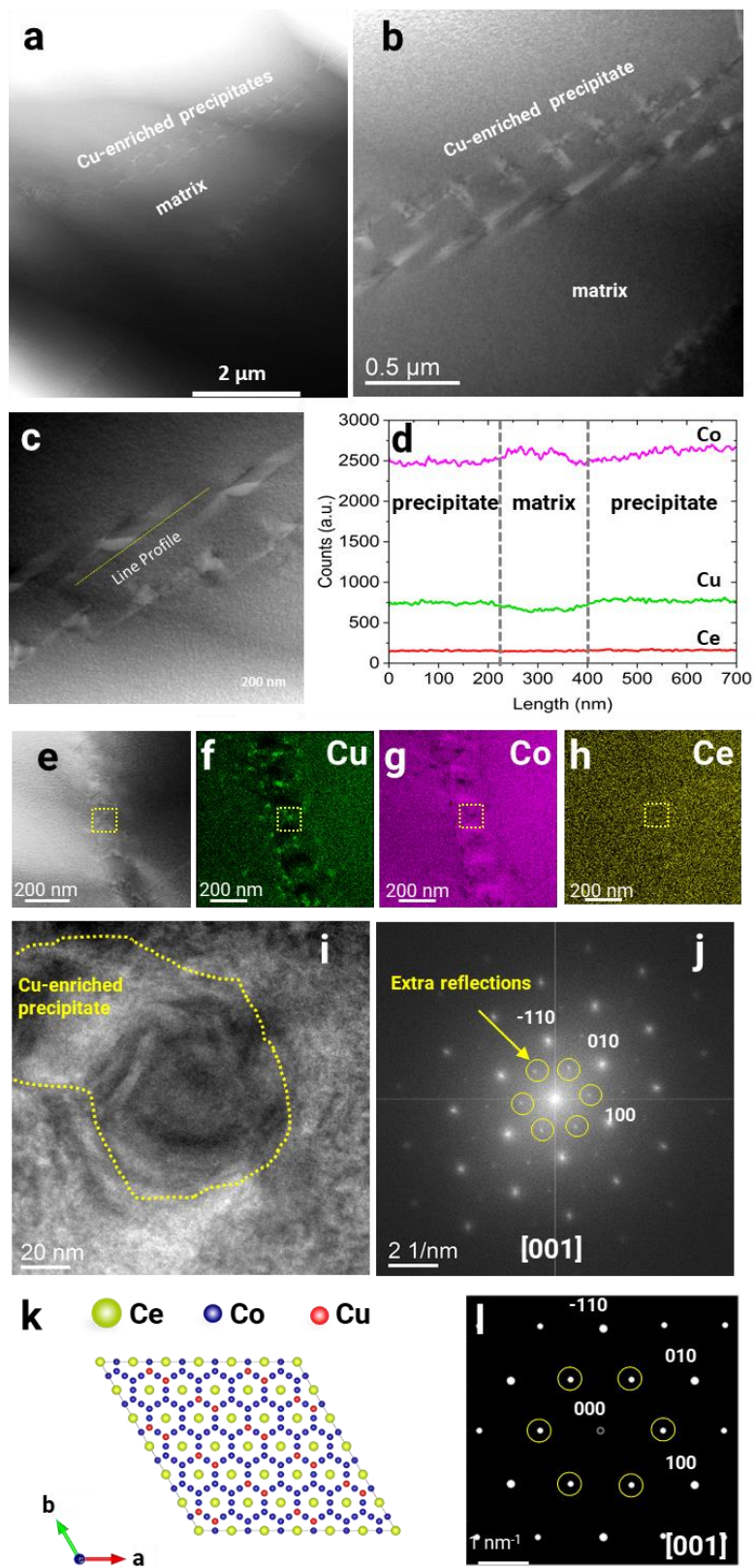

**Figure S9** TEM analysis of a polycrystalline high $H_c$ sample. (**a-c**) Bright-field STEM images with increasing magnification. (**d**) EDX profile across the line indicated in (c). (**e-h**) STEM image and corresponding energy-dispersive X-ray spectroscopy (EDX) chemical maps of Cu, Co and Ce. (**i**) BF-TEM image of a representative precipitate and (**j**) its associated fast Fourier transformed image. (**k**) Structural models with 3×3×1 cells of CeCo$_5$ with Cu-ordering in lines on Co-cites. (**l**) Simulated diffraction along the [001] zone axis using the structural models in (k).

## E. Mesoscale magnetic domain structure

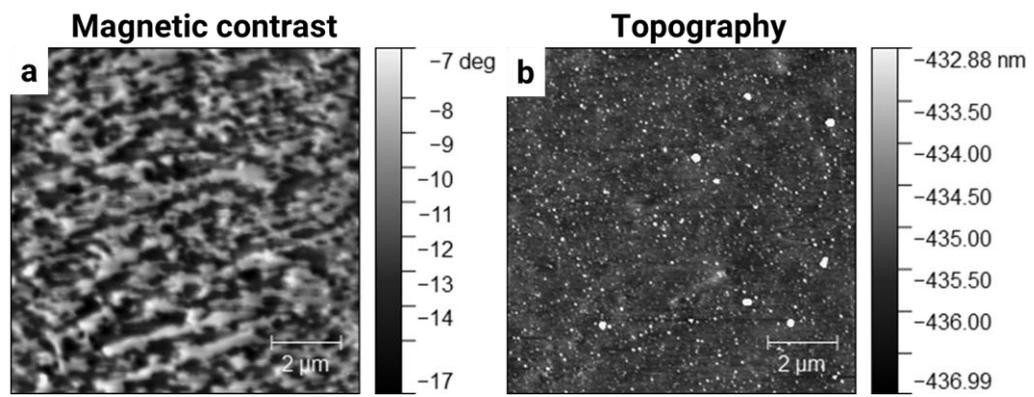

**Figure S10** Magnetic domain structure characterization of the high $H_c$ sample by MFM: (a) Magnetic contrast (b) Topography. The nominal *c*-axis is oriented out of plane.